\documentclass[twocolumn]{aastex6}
\def\kms{km~s$^{-1}$}
\usepackage{graphicx}
\usepackage{amsmath}

\def\dlambda{$\lambda\lambda$}

\slugcomment{Published in {\it The Astrophysical Journal}, 846:50,
  2017 September 1}

\shorttitle{Expos\'e of
  Calcium-rich transients}
\shortauthors{D.\ Milisavljevic et al.}

\begin{document}

\def\cfa{1}
\def\purdue{2}
\def\carnegie{3}
\def\hubble{4}
\def\nw{5}
\def\nyu{6}
\def\ou{7}
\def\sai{8}
\def\osu{9}
\def\iao{10}
\def\uoa{11}
\def\naoj{12}

\title{\MakeLowercase{i}PTF15\MakeLowercase{eqv}: Multi-Wavelength
  Expos\'e of a Peculiar Calcium-Rich Transient}

\author{Dan~Milisavljevic\altaffilmark{\cfa,\purdue},
        Daniel~J.~Patnaude\altaffilmark{\cfa},
        John~C.~Raymond\altaffilmark{\cfa},
        Maria~R.~Drout\altaffilmark{\carnegie,\hubble},
        Raffaella Margutti\altaffilmark{\nw,\nyu},\\
        Atish~Kamble\altaffilmark{\cfa},
        Ryan~Chornock\altaffilmark{\ou},
        James~Guillochon\altaffilmark{\cfa},
        Nathan~E.~Sanders\altaffilmark{\cfa},
        Jerod~T.~Parrent\altaffilmark{\cfa},\\       
        Lorenzo~Lovisari\altaffilmark{\cfa},
        Igor~V.~Chilingarian\altaffilmark{\cfa,\sai},
        Peter~Challis\altaffilmark{\cfa},
        Robert~P.~Kirshner\altaffilmark{\cfa},\\
        Matthew~T.~Penny\altaffilmark{\osu},
        Koichi~Itagaki\altaffilmark{\iao},
        J.~J.~Eldridge\altaffilmark{\uoa},
        Takashi~J.~Moriya\altaffilmark{\naoj}
}

\altaffiltext{\cfa}{Harvard-Smithsonian Center for Astrophysics, 60
  Garden St., Cambridge, MA 02138, USA}
\altaffiltext{\purdue}{Purdue University, Department of Physics and
  Astronomy, 525 Northwestern Avenue, West Lafayette, IN 47907, USA}
\altaffiltext{\carnegie}{The Observatories of the Carnegie Institution
  for Science, 813 Santa Barbara St., Pasadena, CA 91101, USA}
\altaffiltext{\hubble}{Hubble, Carnegie-Dunlap Fellow}
\altaffiltext{\nw}{Center for Interdisciplinary Exploration and
  Research in Astrophysics (CIERA) and Department of Physics and
  Astrophysics, Northwestern University, Evanston, IL 60208, USA}
\altaffiltext{\nyu}{Center for Cosmology and Particle Physics, New
  York University, New York, NY 10003, USA}
\altaffiltext{\ou}{Astrophysical Institute, Department of Physics and
  Astronomy, 251B Clippinger Lab, Ohio University, Athens, OH 45701,
  USA}
\altaffiltext{\sai}{Sternberg Astronomical Institute, M.V.Lomonosov Moscow State University, 13 Universitetsky prospect, Moscow, Russia 119992}
\altaffiltext{\osu}{Department of Astronomy, Ohio State University, 140 West 18th Avenue, Columbus, OH 43210, USA}
\altaffiltext{\iao}{Itagaki Astronomical Observatory, Teppo-cho, Yamagata, Yamagata 990-2492, Japan}
\altaffiltext{\uoa}{Department of Physics, University of Auckland, Private Bag 92019, Auckland, New Zealand}
\altaffiltext{\naoj}{Division of Theoretical Astronomy, National Astronomical Observatory of Japan, National Institutes of Natural Sciences, 2-21-1 Osawa, Mitaka, Tokyo 181-8588, Japan}


\begin{abstract}

  The progenitor systems of the class of ``Ca-rich transients'' is a
  key open issue in time domain astrophysics. These intriguing objects
  exhibit unusually strong calcium line emissions months after
  explosion, fall within an intermediate luminosity range, are often
  found at large projected distances from their host galaxies, and may
  play a vital role in enriching galaxies and the intergalactic
  medium. Here we present multi-wavelength observations of
    iPTF15eqv in NGC 3430, which exhibits a unique combination of
    properties that bridge those observed in Ca-rich transients and
    Type Ib/c supernovae. iPTF15eqv has among the highest
    [\ion{Ca}{2}]/[\ion{O}{1}] emission line ratios observed to date,
    yet is more luminous and decays more slowly than other Ca-rich
    transients. Optical and near-infrared photometry and spectroscopy
  reveal signatures consistent with the supernova explosion of a
  $\la 10$\,M$_{\odot}$ star that was stripped of its H-rich envelope
  via binary interaction. Distinct chemical abundances and ejecta
  kinematics suggest that the core collapse occurred through electron
  capture processes.  Deep limits on possible radio emission made with
  the Jansky Very Large Array imply a clean environment ($n \la$ 0.1
  cm$^{-3}$) within a radius of $\sim 10^{17}$ cm. Chandra X-ray
  Observatory observations rule out alternative scenarios involving
  tidal disruption of a white dwarf by a black hole, for masses $>$
  100 M$_{\odot}$. Our results challenge the notion that
  spectroscopically classified Ca-rich transients only
  originate from white dwarf progenitor systems, complicate the view
  that they are all associated with large ejection velocities, and
  indicate that their chemical abundances may vary widely between
  events.

\end{abstract}

\section{Introduction}

An emergent class of transient objects defined by unusually strong
calcium line emissions that develop in optical spectra months after
explosion has garnered considerable attention in the last
decade. These have been referred to as ``Ca-rich,''
\citep{Filippenko03}, SN\,2005E-like \citep{Perets10}, and ``Ca-rich
gap'' transients \citep{Kasliwal12}. Here we adopt the broad
classification of ``Ca-rich transient.''

During the early photospheric stages, Ca-rich transients have distinct
He lines in their spectra and fall within the supernova (SN) Type Ib
classification (see \citealt{Filippenko97,Gal-Yam16}, and
\citealt{Parrent16} for reviews of the SN classification
scheme). However, Ca-rich transients are less luminous than normal SNe
Ib and reach optically thin (i.e., ``nebular'') stages within two
months post-maximum light. During this stage they exhibit a large
emission line ratio [\ion{Ca}{2}]/[\ion{O}{1}] $>$ 2, which suggests
that the total calcium synthesized is greater than that of normal Type
I supernovae, perhaps by factor of $5-10$ \citep{Perets10}. Estimates
of their rates range from 2\% to 12\% of SNe Ia \citep{Kasliwal12}.

The progenitors of Ca-rich transients are not confidently known. Stars
with initial masses in the range of $8-12\;\rm M_{\odot}$ have been
suggested \citep{Kawabata10,Suh11}, but unlike the majority of
core-collapse SNe, many Ca-rich transients have been hosted by
early-type galaxies lacking obvious massive star
populations. Locations within dwarf galaxies or globular clusters is a
possible resolution to this inconsistency \citep{Yuan13}, but close
examinations of explosion sites have thus far failed to uncover any
sign of in situ star formation
\citep{Perets11,Lyman13,Lyman16,Lunnan17}. \citet{Gvaramadze17}
recently proposed that the Galactic supernova remnant RCW 86 formed
from a core-collapse Ca-rich supernova that polluted the progenitor
star's solar-type binary companion that evolved through a common
envelope phase.

Ca-rich transients are often found at large distances away from their
host galaxies (as high as 150 kpc;
\citealt{Perets10,Kasliwal12,Perets14,Foley15}). This intriguing
property, combined with the lack of nearby stellar populations, has
supported the notion that the progenitor systems have traveled
significant distances before exploding \citep{Lyman14}, perhaps via
interaction with a super-massive black hole (BH) \citep{Foley15}.

The relatively large delay-time distribution required to travel large
distances favors an older white dwarf (WD) population that also
contributes to SNe Ia. Many explosion channels have been proposed,
including helium detonations occurring on a helium-accreting WD
\citep{SB09,Shen10,Perets10,Waldman11,WK11,Dessart15,MH15}, tidal
disruptions of low mass WDs by neutron stars or stellar mass BHs
\citep{Metzger12,Fernandez13,Margalit16}, and tidal detonations of WDs
caused by random flyby encounters with an intermediate mass BH (IMBH)
in dwarf galaxies or globular clusters
\citep{Rosswog08,MacLeod14,Sell15}. Though the class has generally
been treated uniformly, suspicions of heterogeneity in the progenitor
system have been voiced
\citep{Kawabata10,Lyman14,Valenti14,Sell15,Lunnan17}.

Here we present and analyze the multi-wavelength data set of a
supernova exhibiting a revealing mix of properties that bridge Ca-rich
transients and core-collapse SNe. Our observations are provided in
sections \ref{sec:iPTF15eqv} and \ref{sec:observations}, our results
in section \ref{sec:results}, and a detailed comparative analysis of
the entire class of Ca-rich transients in section
\ref{sec:discussion}. In section \ref{sec:conclusions} we review the
ramifications of our conclusion that the progenitor system of the
Ca-rich transient iPTF15eqv was most likely a massive star, and
describe future observations and modeling required to understand the
wide-ranging diversity encompassed in the Ca-rich transient
observational classification.

\begin{figure}[tp]
\centering

\includegraphics[width=\linewidth]{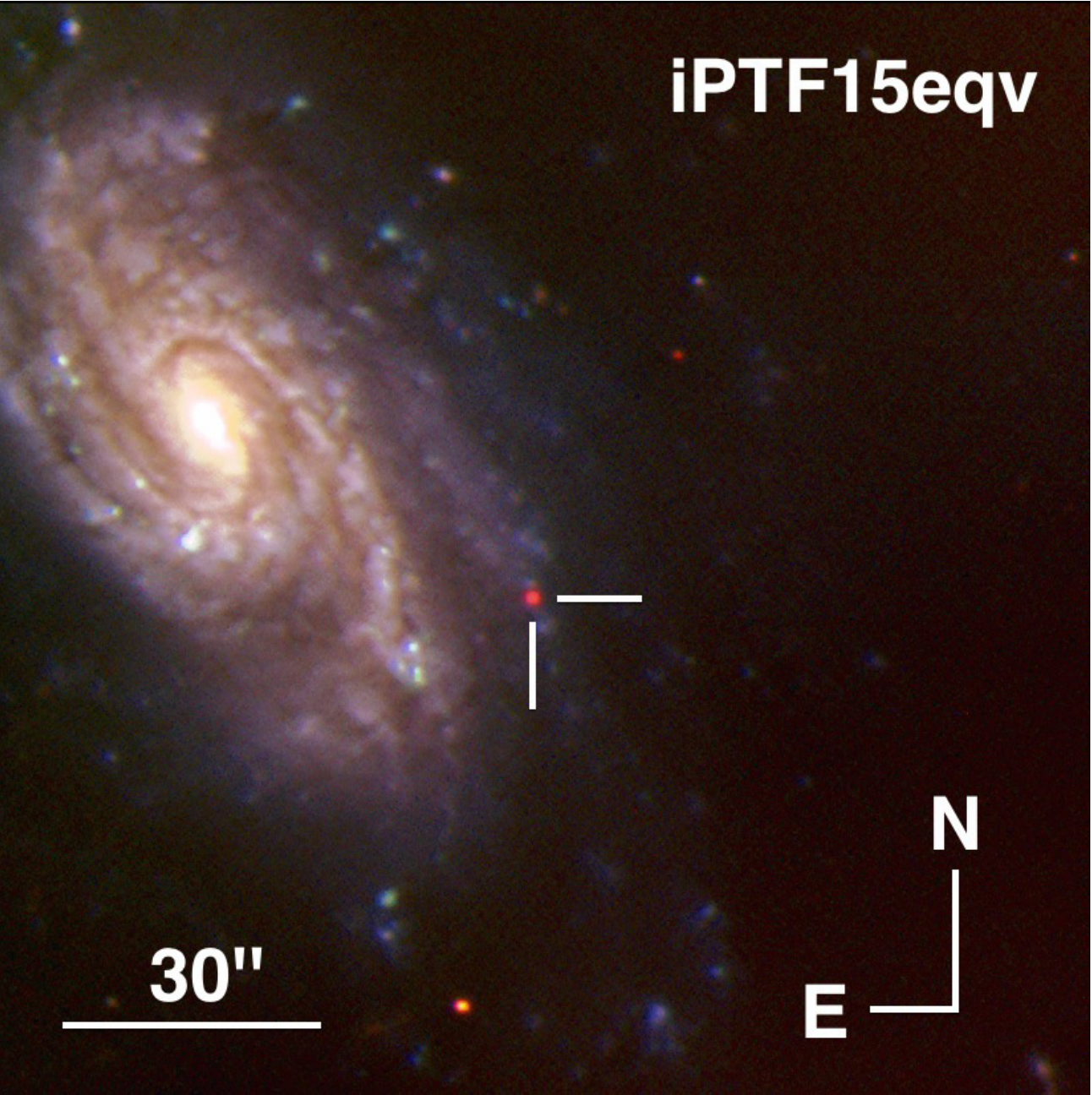}

\caption{Composite image of iPTF15eqv and host galaxy NGC\,3430 made
  from MMT 6.5m telescope + MMTCam observations obtained in
  $g^{\prime}$ (blue), $r^{\prime}$ (green), and $i^{\prime}$ (red)
  bands. The red color of iPTF15eqv reflects the high equivalent width
  of the [\ion{Ca}{2}] $\lambda\lambda$7291, 7324 emission lines.}

\label{fig:fchart}
\end{figure}

\section{\MakeLowercase{i}PTF15\MakeLowercase{eqv}}
\label{sec:iPTF15eqv}

On 2015 September 27.81 (all dates UT) we reported discovery of
PSN\,J10520833+3256394 in NGC 3430 to the Central Bureau for
Astronomical Telegrams ``Transient Object Followup Reports''
page.\footnote{http://www.cbat.eps.harvard.edu/unconf/followups/\\J10520833+3256394.html}
Months later an independent discovery was made by the intermediate
Palomar Transient Factory (iPTF) on 2015 December 6 and designated
iPTF15eqv \citep{Cao15}, which we adopt here. Spectra obtained on 2015
Dec 07.53 by \citet{Cao15} led them to classify iPTF15eqv as a Type
IIb/Ib supernova in the nebular phase.

Description of the spectrum noted weak continuum emission superposed
by strong [\ion{Ca}{2}] $\lambda\lambda$7291, 7324 and weak H$\alpha$
emission lines.  The lack of equally strong [\ion{O}{1}]
$\lambda\lambda$6300, 6364 emission is unusual for a SN IIb/Ib and
prompted our group to obtain optical spectra that confirmed the
non-standard nature of the emissions.\footnote{Y.\ Cao also kindly
  provided the original classification spectrum for us to inspect.} A
multi-wavelength follow up campaign supported by radio observations
with the Karl G.\ Jansky Very Large Array (VLA) and X-ray observations
with Chandra X-ray Observatory (CXO) was initiated.

Distance estimates to the host galaxy NGC 3430 range from 15.4 Mpc
\citep{Theureau07} to 32.9 Mpc \citep{Lagattuta13}. We adopt an
estimate of 30.4 Mpc with distance modulus $\mu = 32.41 \pm 0.43$
\citep{Sorce14}. NGC 3430 is classified as a Type~SBc spiral by
\cite{thirdref}, and also hosted the Type II SN 2004ez
\citep{Nakano04,Filippenko04}.

Extinction toward iPTF15eqv appears to be minimal. Foreground
extinction due to the Milky Way is $E(B-V)_{mw} = 0.021$ mag
\citep{Schlafly11}, there is no obvious suppression of flux at short
wavelengths, and no conspicuous \ion{Na}{1}\,D absorption is observed
in our optical spectra (see section~\ref{sec:spectra} for
details). Thus, no correction for extinction has been adopted in this
paper to the photometry or spectra. The one exception is an estimate
of the metallicity of the host system (section~\ref{sec:metallicity})
that uses the relative strengths of narrow emission lines associated
with the environment around iPTF15eqv.

In Figure~\ref{fig:fchart}, we show a finding chart of iPTF15eqv made
from images of the region obtained with the 6.5-m MMT telescope in
combination with the MMTCam
instrument\footnote{http://www.cfa.harvard.edu/mmti/wfs.html} in
$g^{\prime}r^{\prime}i^{\prime}$-band filters. This composite image
highlights the location of iPTF15eqv, which lies along an exterior
spiral arm approximately 44$^{\prime\prime}$ ($\approx 6.5$~kpc) from
the host galaxy nucleus.

\begin{figure}[tp]
\centering

\includegraphics[width=\linewidth]{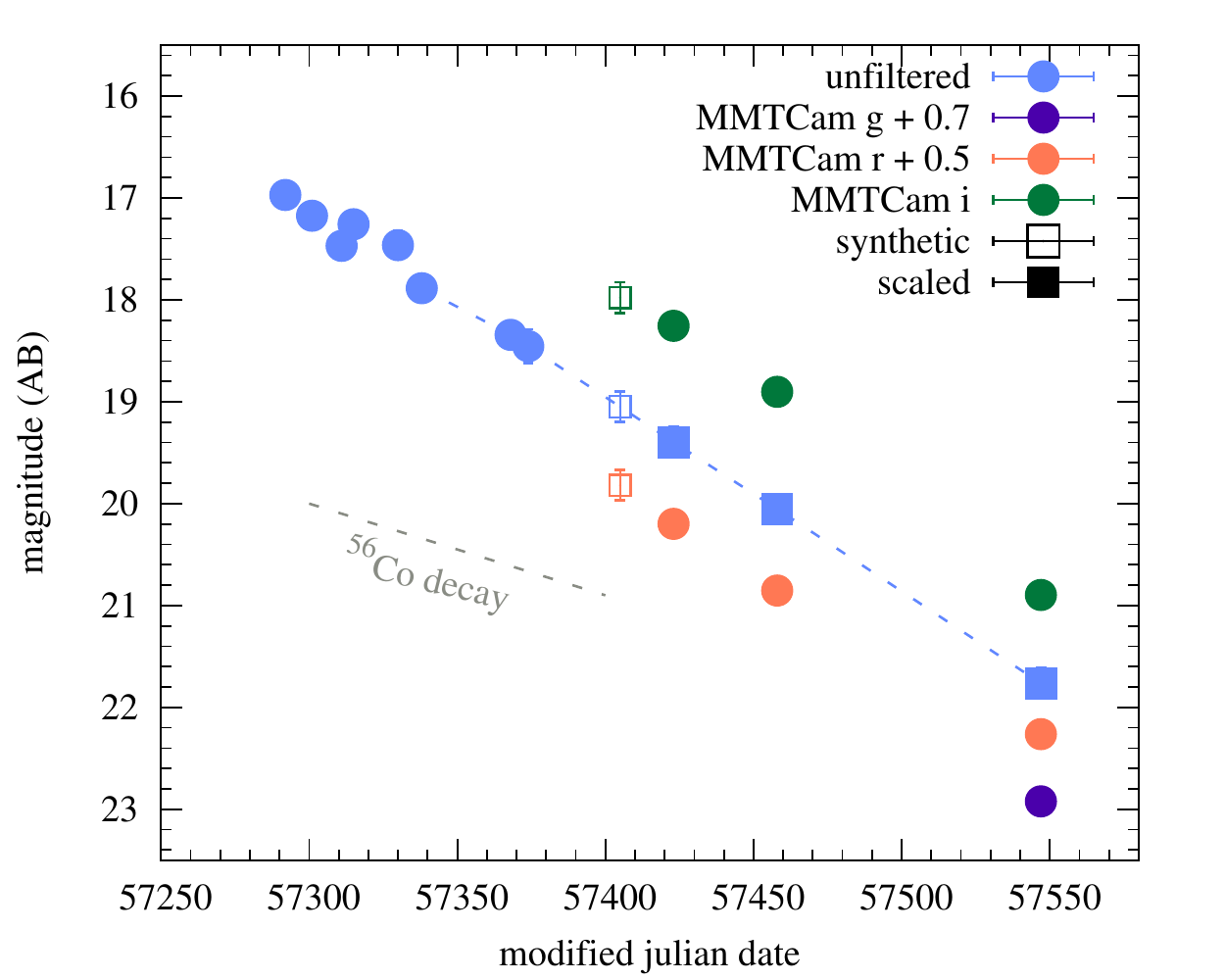}

\caption{Photometry of iPTF15eqv. Unfiltered data are from Itagaki
  Observatory. Sloan $g'r'i'$ photometry is from
  MMTO. Photometric measurements labeled ``synthetic'' are derived
  from optical spectroscopy, while those labeled ``scaled'' are
  derived from Sloan $g'r'$ photometry that has been corrected
  using color terms. Details are provided in the text.}

\label{fig:lc}
\end{figure}

\section{Observations}
\label{sec:observations}

\subsection{Optical photometry}

Unfiltered images of iPTF15eqv and its surrounding environment were
obtained at Itagaki Observatory using the 0.60-m $f/5.7$, 0.50-m
$f/6.0$, and 0.35-m $f/11.0$ telescopes mounted with a Bitran-CCD
(Kodak KAF 1001E) detector\footnote{Specifications of the KAF 1001E
  detector can be found at
  http://www.onsemi.com/pub\_link/Collateral/KAF-1001-D.PDF} between
2015 September 27 and 2015 December 18.  We also obtained three epochs
of MMTCam $r'i'$-band photometry and one epoch of $g'$-band photometry
between 2016 February 06 and 2016 June 08. A log of the observations
is provided in Table~1, and a plot of the final photometry in
Figure~\ref{fig:lc}. Unless otherwise noted, uncertainties are quoted
at the 1$\sigma$ confidence level.
 
Processing of the images was completed using standard routines in the
Image Reduction and Analysis Facility (IRAF).\footnote{IRAF is
  distributed by the National Optical Astronomy Observatories, which
  are operated by the Association of Universities for Research in
  Astronomy, Inc., under cooperative agreement with the National
  Science Foundation.} This included bias, flat-field, and dark frame
corrections, when available. Multiple images in a single night were
stacked. We performed point-spread function (PSF) photometry on
iPTF15eqv and nearby field stars.

Absolute calibration was carried out using Sloan Digital Sky Survey
(SDSS) observations of field stars. Approximate unfiltered magnitudes
for the field stars were calculated by performing synthetic photometry
on the SDSS $u'g'r'i'z'$ spectral energy distribution using the
typical spectral response achieved by the KAF 1001E detector in use at
Itagaki Observatory.  In addition, we subtract the flux associated
with an $\sim$21 mag stellar knot located 1.5$^{\prime\prime}$ from
iPTF15eqv from the unfiltered magnitudes measured from the Itagaki
Observatory images.  This stellar knot is well resolved and separated
from the SN in our MMTCam observations, but would be included within
the aperture used on the Itagaki Observatory images (typical
resolution $\approx$4$^{\prime\prime}$). This correction is
$\lesssim$0.1 mag for all epochs.
  
In order to directly compare the early unfiltered light curve to our
late-time MMTCam observations, we performed synthetic photometry on a
spectrum of iPTF15eqv obtained on 2016 January 18 (see
section~\ref{sec:spectra}) to investigate the relationship between the
unfiltered magnitudes of the KAF 1001E detector and standard
bandpasses.  We verified the relationship using spectra of the Ca-rich
transient SN\,2005E \citep{Perets10}. For transients with large
$r'$-$i'$ colors (as is observed for iPTF15eqv at late times due to
the large [\ion{Ca}{2}] feature) we find an approximate relationship
between unfiltered and $r'$-band magnitudes which ranges from $r'-$unf
= 0.0 mag for $r'-i'$ $= 0.85$ mag to $r'-$unf $= 0.3$ mag for $r'-i'$
$= 1.45$ mag.

Applying these shifts to our MMTCam data we obtain approximate
unfiltered magnitudes for iPTF15eqv at late epochs that can be
compared to the earlier phase data from Itagaki Observatory.  These
data are shown as open squares in Figure~\ref{fig:lc}. We also plot
unfiltered, $r'$- and $i'$-band magnitudes obtained from our synthetic
photometry. We do not attempt to transform the unfiltered magnitudes
obtained from the Itagaki Observatory images to a standard bandpass
since no spectral or color information is available at these epochs.

\begin{figure}[tp]
\centering

\includegraphics[width=\linewidth]{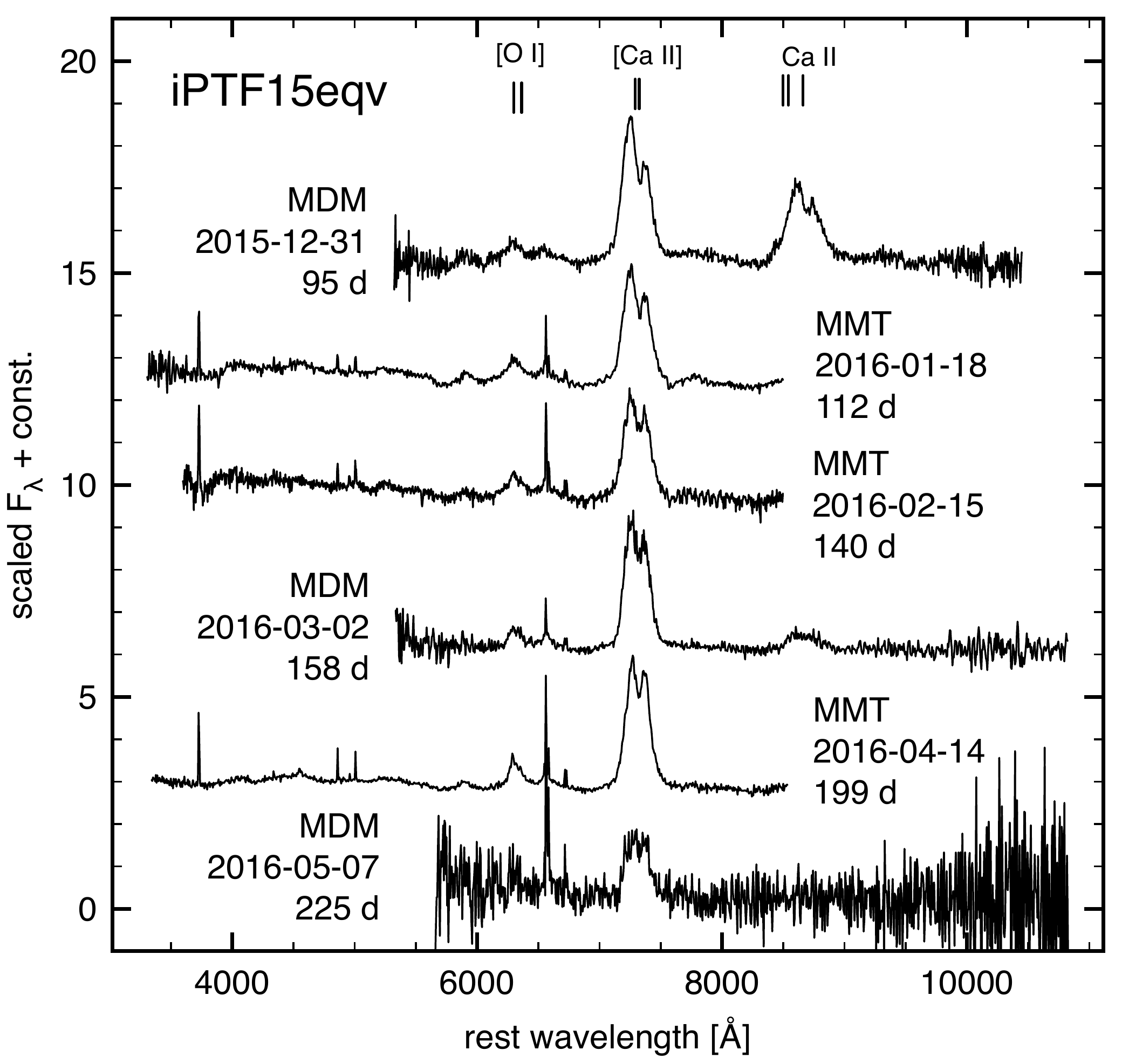}

\caption{Six epochs of optical spectra of iPTF15eqv. Telescopes used,
  dates of observations, and phases with respect to the date of
  discovery on 2015 September 27.8 are provided. The three strongest
  emission lines, [\ion{Ca}{2}] $\lambda\lambda$7291, 7324, the
  \ion{Ca}{2} triplet, and [\ion{O}{1}] $\lambda\lambda$6300, 6364,
  are identified.}

\label{fig:spectra}
\end{figure}

\subsection{Optical spectroscopy}
\label{sec:spectra}

Low-resolution optical spectra of iPTF15eqv were obtained using the
MMT telescope in combination with the Blue Channel instrument
\citep{Schmidt89}, and the Hiltner 2.4\,m telescope in combination
with the Ohio State Multi-Object Spectrograph (OSMOS;
\citealt{osmospaper}) instrument and R4K detector at MDM
observatory. Conditions were photometric for most
observations. Standard stars from \citet{Oke90} and \citet{Hamuy92}
were observed for spectrophotometric calibration and to remove
telluric features. A log of all optical spectroscopy is provided in
Table~2.

The Blue Channel observations used the B300 grating in combination
with a 1$^{\prime\prime}$ slit width yielding spectra spanning a
wavelength window of $3400 - 8500$~\AA\ and with resolution
$R \sim 740$. OSMOS observations used a 1.2\arcsec\ slit width with
the VPH-red grism and an OG530 order-blocking filter, resulting in
spectra spanning a wavelength window of approximately
$5500-10500$~\AA\ with resolution $R \sim 1200$.

Standard procedures to bias-correct, flat-field, and flux calibrate
the data were followed using the IRAF/PYRAF software\footnote{PYRAF is
a product of the Space Telescope Science Institute, which is operated
by AURA for NASA.} and our own IDL routines. Narrow and unresolved
emission lines of [\ion{O}{2}] 3727, [\ion{O}{3}]
4959, 5007, H$\alpha$, H$\beta$, and [\ion{S}{2}] 6716, 6731
attributable to a coincident \ion{H}{2} region were used to determine
a recession velocity of 1494 \kms, which was removed from all
spectra. A plot of the observations labeling prominent emission
features is presented in Figure~\ref{fig:spectra}.

\begin{figure}[tp]
\centering

\includegraphics[width=\linewidth]{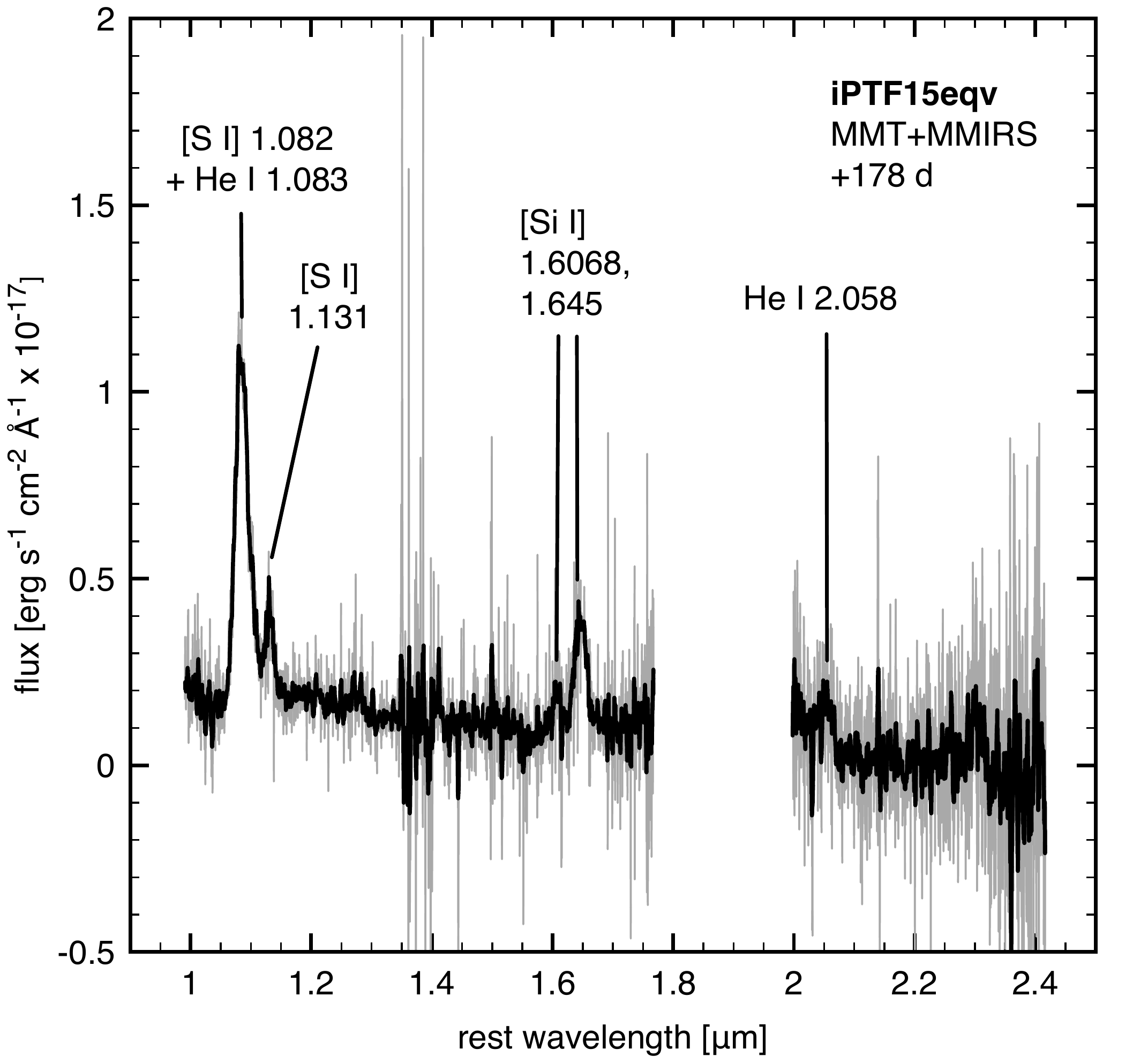}

\caption{NIR spectrum of iPTF15eqv obtained on 2016 March 23 using the
  6.5m MMT telescope and MMIRS instrument. Line identifications are
  provided for the strongest emissions.}

\label{fig:nirspec}
\end{figure}

\subsection{Near-infrared spectroscopy and photometry}

\begin{figure*}[tp]
\centering

\includegraphics[width=0.49\linewidth]{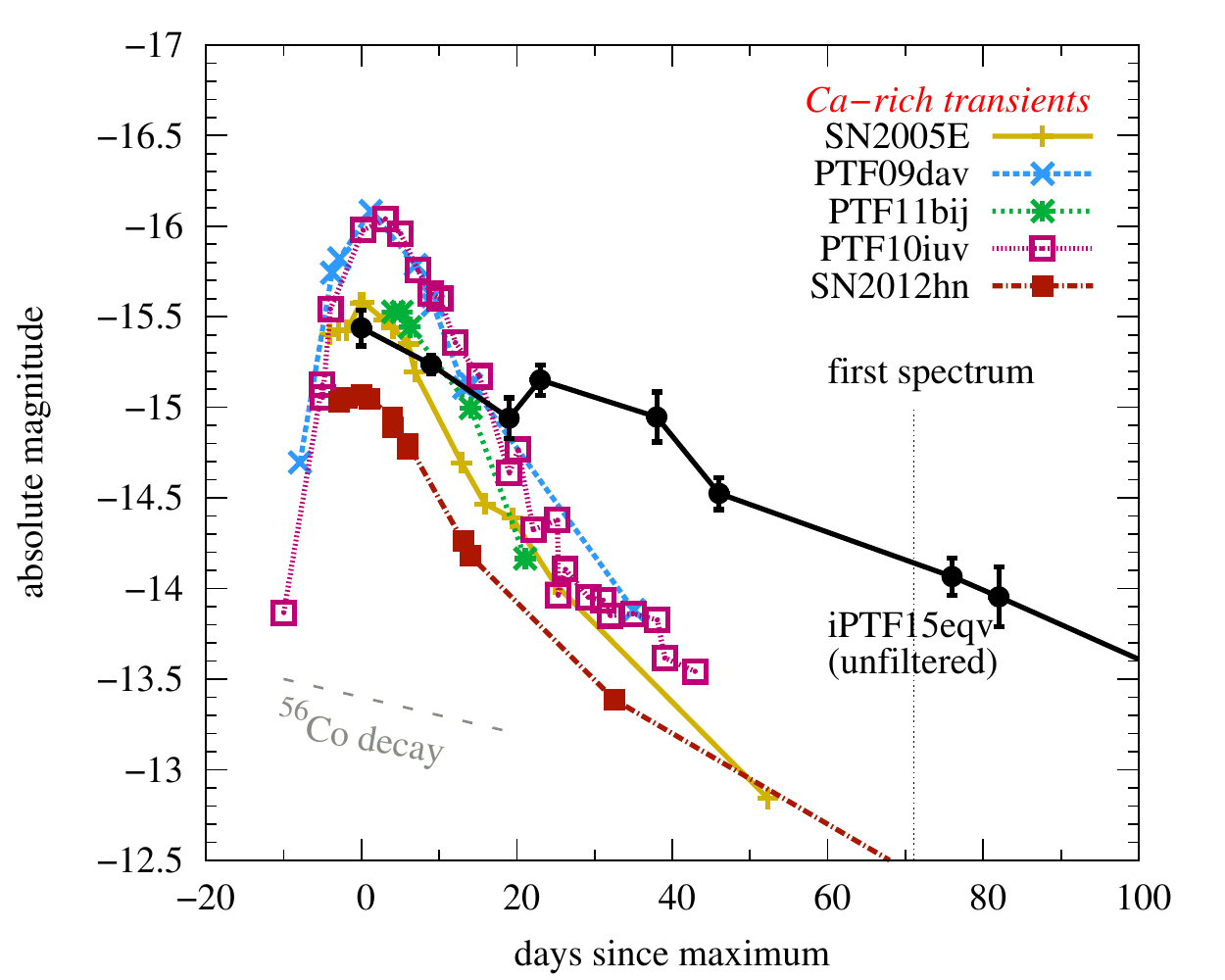}
\includegraphics[width=0.49\linewidth]{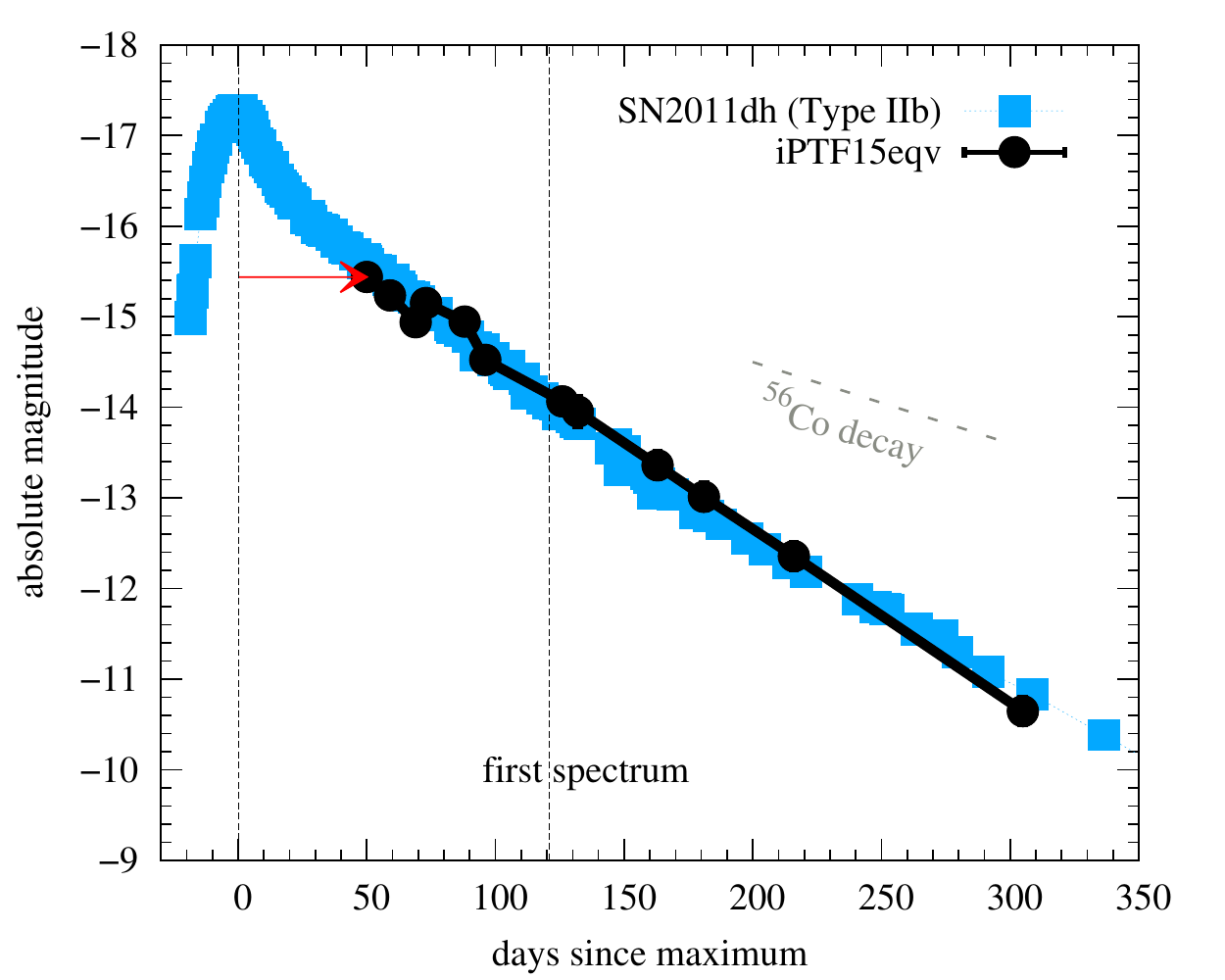}

\caption{Comparing the unfiltered light curve of iPTF15eqv to Ca-rich
  transients and the Type IIb SN\,2011dh. {\it Left}: The absolute
  $R$-band light curves of Ca-rich transients. Here we assume that
  iPTF15eqv was discovered around maximum light. Because the Itagaki
  Observatory photometry is a non-standard combination of $R$- and
  $I$-band, the light curve of iPTF15eqv shown may deviate from its
  true $R$-band-only light curve by $\la 0.3$ mag. Data are from
  \citet{Kasliwal12} and \citet{Valenti14}. {\it Right}: The $R$-band
  light curve of the Type IIb SN\,2011dh. Data are from
  \citet{Ergon15}. Here we assume that the time of maximum light for
  iPTF15eqv was 50 days previous to the time of discovery. A distance
  modulus of $\mu = 32.41$ has been used and no extinction has been
  corrected. }

\label{fig:ca-rich-lc}
\end{figure*}

A single epoch of low-resolution, near-infrared (NIR) spectra spanning
$1-2.4\,\mu$m was obtained on 2016 March 23 with MMT using the MMT and
Magellan Infrared Spectrograph (MMIRS) instrument
\citep{McLeod12}. Observations were made using a 0.6$^{\prime\prime}$
slit width in three configurations: zJ filter (spanning 0.95 - 1.50
$\mu$m) + J grism ($R \sim 2000$), H filter (spanning 1.50-1.79
$\mu$m) + H3000 grism ($R \sim 3000$), and Kspec filter (spanning
1.95-2.45$\mu$m) + K3000 grism ($R \sim 3000$).  Exposure times were
$4\times 300$\,sec for each configuration, and the slit was dithered
between exposures.  Data were processed using the standard pipeline
\citep{Chilingarian15}, to develop wavelength calibrated 2D frames
from which 1D extractions were made. A plot of these data combining
all three individual spectra is shown in
Figure~\ref{fig:nirspec}. Line identifications are discussed in
section \ref{sec:identify}.

Acquisition images obtained in J-band were used to flux calibrate our
NIR spectra. We performed aperture photometry on a stacked image
made from pipeline-processed individual images, choosing an aperture
to encompass all of the SN flux.  2MASS magnitudes of field stars
provided a means of absolute calibration. We estimate the J-band
magnitude of iPTF15eqv on 2016 March 23 to be $19.77 \pm 0.16$ mag.

\subsection{Chandra X-ray observations}
\label{sec:x-ray}

We obtained deep X-ray observations of iPTF15eqv with CXO on 2016
January 4 under our approved program to explore stellar deaths at high
energies and radio wavelengths (observation ID 16821, PI:
Margutti). CXO data have been reduced with the CIAO software v4.8 and
corresponding calibration files. No X-ray source is detected at the
location of iPTF15eqv with a total exposure time of 9.9 ks. The
inferred $3\,\sigma$ count-rate limit is
$4\times 10^{-4}\,\rm{c\,s^{-1}}$ (0.5-8 kev). The neutral hydrogen
column density in the direction of iPTF15eqv is
$1.84\times 10^{20}\,\rm{cm^{-2}}$ \citep{Kalberla05}. For an assumed
non-thermal spectrum with index $\Gamma=2$, we derive an unabsorbed
(absorbed) flux limit of $4.8\times 10^{-15}\,\rm{erg\,s^{-1}cm^{-2}}$
($4.5\times 10^{-15}\,\rm{erg\,s^{-1}cm^{-2}}$) in the 0.3-10 keV
energy range, corresponding to $5.3\times 10^{38}\,\rm{erg\,s^{-1}}$
at the distance of iPTF15eqv.

\subsection{Jansky Very Large Array radio observations}

We obtained five epochs of radio observations of iPTF15eqv at
frequencies ranging from 4.9 to 16.0 GHz with the VLA (PI: Margutti,
program SG0534). All observations were taken in the standard continuum
observing mode with a bandwidth of
$\rm 16 ~IF \times 64 ~channel/IF \times 2 ~MHz/channel$. The VLA
changed configurations at various stages during these
observations. During the reduction we split the data in two side bands
($\rm 8 ~IF$ each) of approximately 1 GHz each. We used radio source
3C286 for flux calibration, and calibrator J1018+3542 for phase
referencing. Data were reduced using standard packages within the
Astronomical Image Processing System (AIPS). No radio emission was
detected from iPTF15eqv in any of these observations resulting in deep
flux density limits. The results have been summarized in Table~3.

\section{Results}
\label{sec:results}

\subsection{Light curve evolution}
\label{sec:lc}

We have utilized all photometric data of iPTF15eqv to compare it to
the light curves of selected Ca-rich transients. Although our
unfiltered photometry does not strictly match $R$-band photometry, it
does provide a reasonable means of comparing the slope and absolute
luminosity of the light curve.  As shown in Figure
\ref{fig:ca-rich-lc}, iPTF15eqv was discovered at its peak observed
unfiltered absolute magnitude of $-15.4$, which is in the range of
Ca-rich objects ($R \sim -15$ to $-16.5$ mag;
\citealt{Kasliwal12,Valenti14}).  The decline of the light curve of
iPTF15eqv at a rate of $\approx 0.017$\,mag\,day$^{-1}$ is shallower
than those of other Ca-rich transients at similar epochs. If we assume
that iPTF15eqv was discovered around maximum light and shares the
typical rise time of Ca-rich transients (12$-$15 days;
\citealt{Kasliwal12}), then the explosion date was approximately 2015
September 12. We refer to this as $t_{\rm exp}(s1)$, and consider it a
plausible scenario since the first reported spectrum of iPTF15eqv
showing strong [\ion{Ca}{2}] emissions by \citet{Cao15} was obtained
70 days after discovery, which is within the time frame that Ca-rich
transients exhibit pronounced Ca emissions.

However, because we do not have firm observational constraints on the
explosion date, the light curve of iPTF15eqv may have peaked many
weeks or months before the discovery date. To investigate this
alternative scenario, we attempted to match the light curve of
iPTF15eqv to those of SN Ib/c \citep{Elmhamdi06,Drout11}. By shifting
the light curve of iPTF15eqv $\approx 50$ days relative to the date of
discovery, the overall slope and luminosity is comparable to the
late-time linear decay rates found in other SN Ib/c and consistent
with expectations from $^{56}$Co-decay (Figure
\ref{fig:ca-rich-lc}). A particularly good match was found with the
light curve of SN\,2011dh (\citealt{Ergon15}). If this analogous
connection to SN\,2011dh is valid, then the first spectrum of
iPTF15eqv was obtained some 120 days after the light curve peak that
may have have been as luminous as $-17.5$ mag. We refer to the
explosion date 2015 July 24 of this alternative scenario as
$t_{\rm exp}(s2)$. iPTF15eqv most likely exploded in the time interval
$t_{\rm exp}(s1) > t > t_{\rm exp}(s2)$, and peaked in $R$-band
absolute magnitude between approximately $-17.5$ and $-15$.

If iPTF15eqv was, indeed, discovered after maximum light, then its
light curve evolution can be approximately described by the nebular
light curve model of \citet{Valenti08a}, which is based on the
instantaneous rate of energy deposition from the $^{56}$Ni
$\rightarrow$ $^{56}$Co $\rightarrow$ $^{56}$Fe decay chain
\citep{Arnett82}. We investigated the range of viable explosion
parameters for iPTF15eqv in this scenario.  Model parameters include
the total nickel mass, M$_{\rm{Ni}}$, and a constant, $F$, which
describes the incomplete trapping of gamma-rays. $F$ is a function of
the total ejecta mass, M$_{\rm ej}$, explosion kinetic energy,
E$_{\rm{K}}$, and ejecta density distribution, and we adopt the
parameterization of \citet{Valenti08a} such that $F \propto$
M$_{\rm{ej}}$/$\sqrt{E_{\rm{K}}}$ (also see \citealt{Clocchiatti97}).
The luminosity of iPTF15eqv is estimated from the unfiltered light
curve, assuming zero bolometric correction.

\begin{figure}[tp]
\centering
\includegraphics[width=\linewidth]{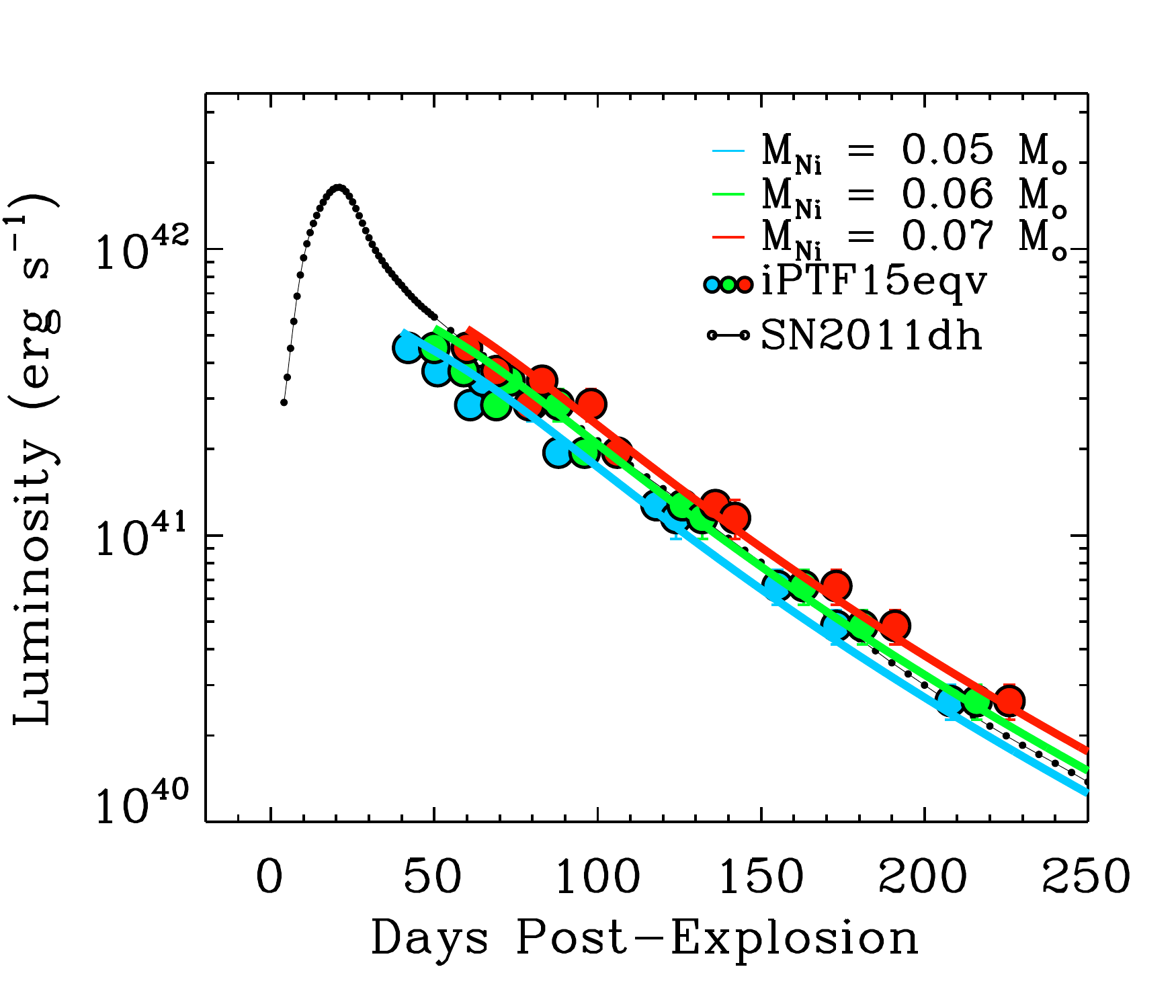}

\caption{Estimating the explosion parameters M$_{\rm Ni}$, E$_{k}$,
  and M$_{\rm ej}$ of iPTF15eqv assuming an explosion date around
  $t_{\rm exp}(s2)$. Colored circles show the light curve of iPTF15eqv
  assuming the explosion occurred 40, 50, and 60 days prior to the
  discovery date of 2015 September 27 (cyan, green, and red,
  respectively). Colored lines show model fits to these light curves
  with 0.05, 0.06 and 0.07 M$_\odot$ of $^{56}$ Ni, respectively. The
  bolometric light curve of SN\,2011dh \citep{Ergon15} is shown for
  reference (black).}

\label{fig:explosionparams}
\end{figure}

The range in mass of $^{56}$Ni required to reproduce the luminosity of
iPTF15eqv is degenerate with an adopted explosion date. The slope of
the light curve rules out reasonable fits bracketing
$t_{\rm exp}(s1)$. However, for explosions bracketing
$t_{\rm exp}(s2)$ (i.e., 40 $-$ 60 days prior to discovery), we obtain
reasonable fits and constrain M$_{\rm{Ni}}$ to be between
$\approx 0.04-0.07$ M$_{\odot}$ (Figure~\ref{fig:explosionparams}). In
order to match the observed decay rate with this model, we require
that (M$_{\rm{ej}}$/M$_\odot$)(E$_{\rm{K}}$/10$^{51}$ erg)$^{-0.5}$
$=$ 3.0 $\pm$ 0.1.  Assuming an average bulk explosion velocity at
peak for Ca-rich transients of $(7-9) \times 10^3$\,\kms\
\citep{Kasliwal12}, we estimate M$_{\rm{ej}} \approx 2-4$ M$_{\odot}$ and
E$_{\rm{K}} \approx (0.8-2) \times 10^{51}$\,erg. These explosion
parameters are comparable to those found for stripped-envelope SNe,
including SN\,2011dh \citep{Bersten12}.

\begin{figure}[tp]
\centering

\includegraphics[width=\linewidth]{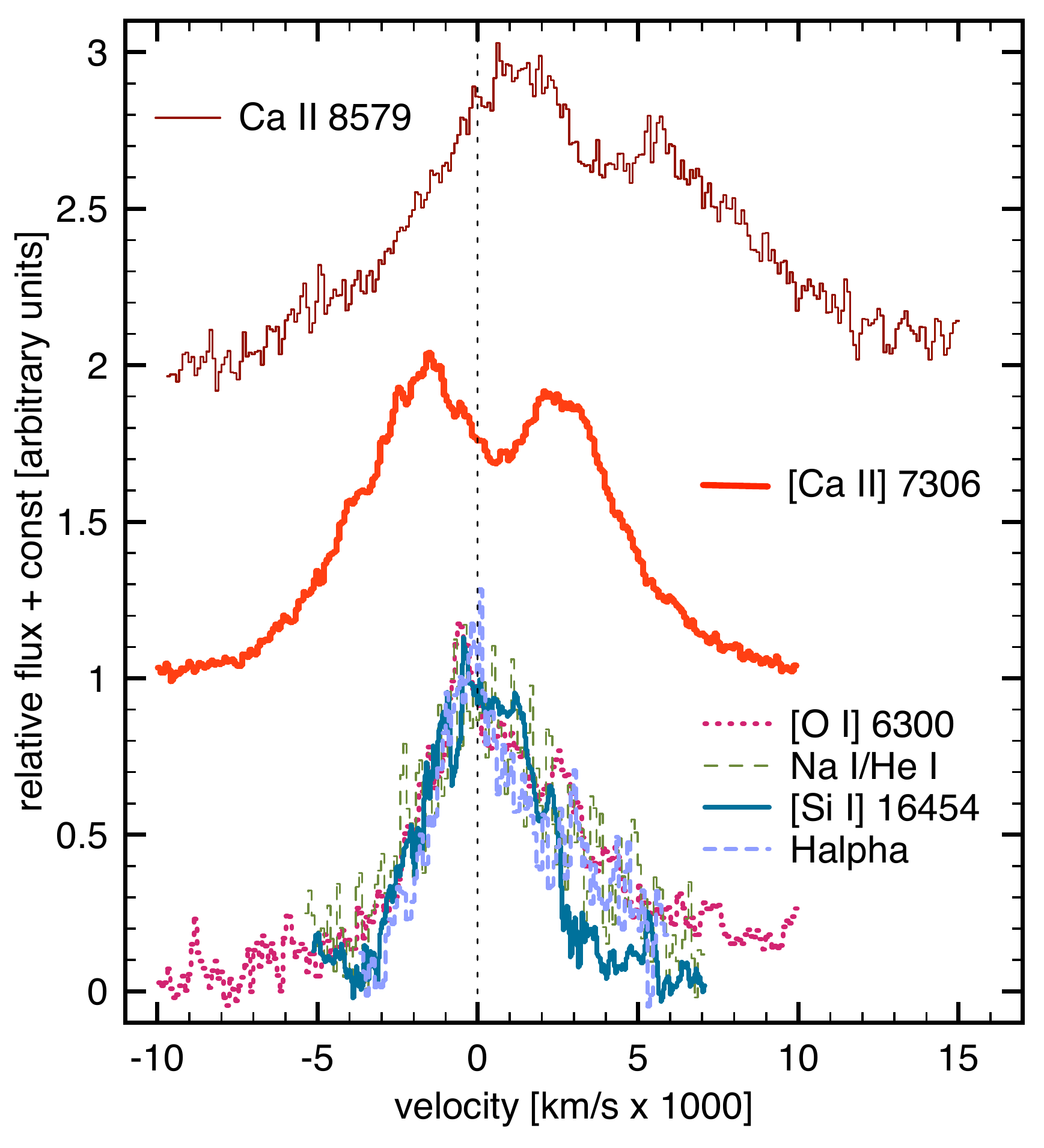}

\caption{Emission line profiles of iPTF15eqv enlarged around the
  \ion{Ca}{2} near-infrared triplet, [\ion{Ca}{2}] 7291, 7324,
  [\ion{O}{1}] 6300, 6364, [\ion{Si}{1}] 16454, and
  H$\alpha$. Velocities are with respect to 8579~\AA, 7306~\AA,
  6300~\AA, 16454~\AA, and 6563 \AA, respectively. From the H$\alpha$
  profile we have removed strong and unresolved H$\alpha$ and
  [\ion{N}{2}] lines from a coincident \ion{H}{2} region, and a
  linearly sloped continuum introduced from blending with the nearby
  [\ion{O}{1}] profile. See section \ref{sec:identify} for discussion
  of why the H$\alpha$ emission feature may have significant
  contribution from \ion{Ca}{1}] 6572.}

\label{fig:lineprofiles}
\end{figure}

\subsection{Emission line identification, Doppler velocities, and
  evolution}
\label{sec:identify}

The dominant emission feature of iPTF15eqv over all observed epochs is
[\ion{Ca}{2}] \dlambda 7291, 7324 (Figure~\ref{fig:spectra}). The
\ion{Ca}{2} NIR triplet \dlambda 8498, 8542, and 8662 is strong on day
95, but has faded by day 158. The critical density for the
[\ion{Ca}{2}] doublet is $\sim 3 \times 10^6~\rm cm^{-3}$ from CHIANTI
v7 \citep{Landi13}, so this threshold may have been reached during the
observed epochs, but the ratio also depends on temperature.

Also unambiguously detected is [\ion{O}{1}] $\lambda\lambda$6300,
6364. The ratio of [\ion{Ca}{2}]/[\ion{O}{1}] $\approx 10$ exhibits
little change over the entire observing period. We attribute emission
centered around 5890~\AA\ to a blend of \ion{Na}{1} and \ion{He}{1}
5876. The central wavelength of the observed distribution favors
\ion{Na}{1} as the dominant ion, but helium is inferred to be a
potential contributor because Ca-rich transients show helium in early
spectra and there is evidence of the \ion{He}{1} lines in the NIR.

We identify a feature centered near 6560~\AA\ as H$\alpha$ or
\ion{Ca}{1}] $\lambda$6572. The central wavelength of the feature
favors H$\alpha$. However, the lack of H$\beta$ emission, the
extremely strong [\ion{Ca}{2}] lines, the large \ion{Ca}{1}] 6572
collision strength \citep{SB01} and the presence of [\ion{Si}{1}] and
[\ion{S}{1}] in the NIR spectrum all suggest that \ion{Ca}{1}] 6572
may be strong.  [\ion{N}{2}] $\lambda\lambda$6548, 6583 may also
contribute to the profile \citep{Jerkstrand15}.

In the day 199 spectrum a minor emission peak is observed around
4550~\AA. We attribute this to blueshifted \ion{Mg}{1}] 4571. An
alternative identification of \ion{S}{1} $\lambda$4589 seems unlikely
since its stronger companion line $\lambda$7725 is not detected.  Also
seen in the day 199 spectrum is a ``plateau'' of emission between
$4000-5600$\,\AA. This emission is not uncommon in SNe Ib/c (see,
e.g., \citealt{Milisavljevic13-12au}), and due largely to iron-peak
elements.

We show the emission line profile distributions for a variety of ions
in Figure~\ref{fig:lineprofiles}. The [\ion{Ca}{2}] lines exhibit the
largest velocities. With respect to the $\lambda$7291 line measuring
toward short wavelengths, and the $\lambda$7324 line measuring toward
long wavelengths, the emission is above the continuum level for
velocities up to and potentially exceeding $\sim 10^{4}$~\kms. The
profile has a central absorption notch centered around $+750$ \kms\
situated between two peaks at velocities of $-1700$ \kms\ and $+2600$
\kms. The blue/red ratio between peaks decreases from 3:1 to 3:2 over
the epochs observed. The central dip in [\ion{Ca}{2}] 7291, 7324 does
not correspond to the separation of the lines (1400 $\rm km~s^{-1}$)
and must be due to geometry or optical depth.  On the other hand, the
Ca triplet on day 95 shows a similar dip, but it is redshifted by
$\approx 4000$ \kms\ and the peaks correspond to the separation
between the two strongest lines of the triplet.  The triplet is almost
certainly optically thick, and radiative transfer must play an
important role.

The [\ion{O}{1}], \ion{Na}{1}/\ion{He}{1} blend, and
H$\alpha$/\ion{Ca}{1}]/[\ion{N}{2}] blend all have velocity spans much
less than that of [\ion{Ca}{2}] and do not share the double-peaked
distribution. The measurement is complicated by a non-linear continuum
and blending of features, but generally for all emission lines except
for calcium we measure maximum velocities in the region of
$\approx 3400$ \kms. Nonuniform density, uneven $^{56}$Co heating, and
chemical stratification are all possible reasons for the observed line
asymmetry. An important consideration is that Ca is relatively easy to
ionize (Ionization Potential = 6.11 eV) compared to Si (IP=8.51 eV)
and S (IP = 10.34 eV), and that the neutral gas may be expanding more
slowly.  In principle if the radioactive material were evenly mixed,
the heating would be proportional to $n$ and the cooling (if optically
thin) proportional to $n^2$, so denser regions would be cooler and
more neutral.

The emission feature centered near 1.64\,$\mu$m could potentially be
ascribed to [\ion{Fe}{2}], but the absence of the stronger
[\ion{Fe}{2}] line at 1.257 $\mu$m makes this impossible.  Instead,
the 1.64\,$\mu$m line must be [Si I] 1.645\,$\mu$m, an identification
confirmed by the 1.6068\,$\mu$m [Si I] line with the intrinsic 1:3
intensity ratio for this pair of lines.

\begin{figure}[tp]
\centering

\includegraphics[width=\linewidth]{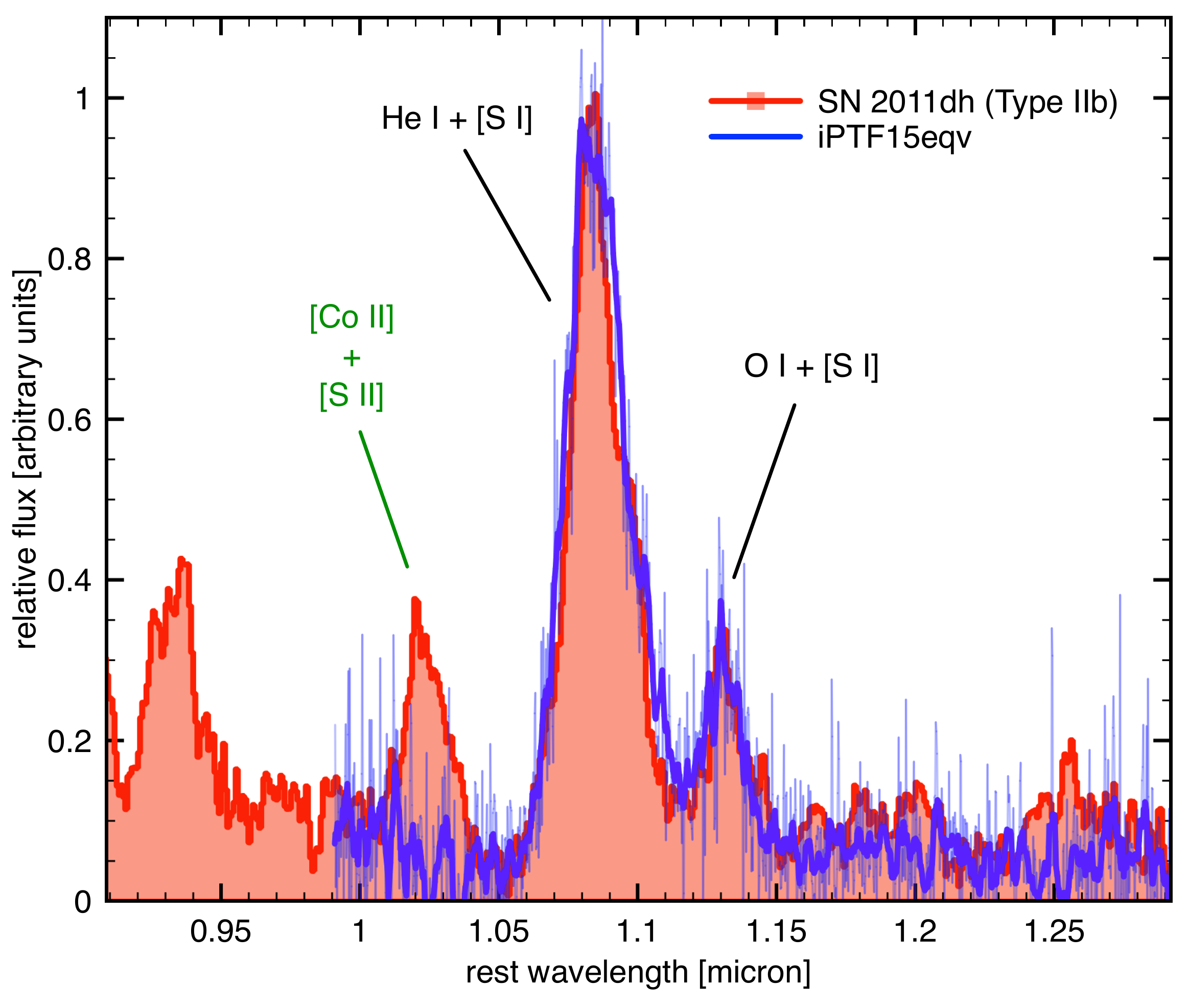}

\caption{NIR spectra of iPTF15eqv (this paper) and SN\,2011dh
  \citep{Ergon15} in the region of 0.95 - 1.25 $\mu$m. Emission peaks
  at 1.083 and 1.125 $\mu$m are common to both, but only SN\,2011dh
  has an emission peak at 1.03 $\mu$m. }

\label{fig:NIRenlarge}
\end{figure}

The presence of the [\ion{Si}{1}] line pair suggests that the lines at
1.08 and 1.13 $\mu$m could be the corresponding [\ion{S}{1}] 1.0824
and 1.1309~$\mu$m lines, rather than the \ion{He}{1} 1.083 and O I
1.129~$\mu$m lines near these wavelengths (e.g.,
\citealt{Jerkstrand15}).  The absence of the \ion{O}{1} 1.317~$\mu$m
line and the \ion{O}{1} multiplet at 7774~\AA, which should be 30 
times stronger than the 1.129 $\mu$m line, rule out the \ion{O}{1}
identification and place an upper limit of 10\% to the \ion{O}{1}
contribution to the 1.13 $\mu$m feature.

The measured intensity ratio $I$(1.08)/$I$(1.13) is 4.2:1, rather than
the 3.4:1 ratio given by the [\ion{S}{1}] Einstein A values,
suggesting that \ion{He}{1} may indeed contribute to the 1.08~$\mu$m
line.  [\ion{Si}{1}] 1.099 may also contribute to the feature
\citep{Fransson89,Jerkstrand15}. A feature consistent with \ion{He}{1}
2.058~$\mu$m, which should be a few times fainter than the
1.083~$\mu$m line, is marginally detected (Figure~\ref{fig:nirspec}).
The optical spectra indicate an upper limit to the \ion{He}{1} 5876
line of $1.3 \times 10^{-15}~\rm erg~cm^{-2}~s^{-1}$, which would
indicate an upper limit of 12\% contribution of He I to the 1.08
$\mu$m feature.  This is within the measurement uncertainties of the
value of 20\% indicated by the difference between the observed line
ratio and the intrinsic [S I] intensity ratio.  We note that the
[\ion{Si}{1}] and [\ion{S}{1}] lines have very low excitation
thresholds, so that they can be excited in fairly cool gas, while the
He I and O I lines are produced by recombination and require ionized
helium and oxygen.  The absence of [\ion{S}{2}] 1.03 $\mu$m emission
while [\ion{S}{1}] emission is strong (Figure~\ref{fig:NIRenlarge}),
implies either predominantly neutral sulfur or very cool gas.

\subsection{Chemical abundances}
\label{sec:abundances}

We estimated chemical abundances of the ejecta of iPTF15eqv using the
relative line strengths and luminosities of the ions identified in our
optical and NIR spectra. The ratio of the Ca triplet to the [\ion{Ca}{2}]
lines at 7300 \AA\/ depends on temperature and density.  On day 95,
the ratio is 0.7, while on day 158 it has fallen to 0.1.  The Ca
triplet lines are produced by decay of the 4p level, which decays by
way of the H and K lines at 3933, 3968 \AA\/ 94\% of the time.
However, for large optical depths in H and K, nearly all the H and K
photons will be converted to triplet photons, which escape because of
the smaller optical depth.

Using atomic rates from CHIANTI version 8 \citep{DelZanna15} and
assuming that all excitations to the 4p level produce Ca triplet
photons, we compute the [Ca II]/Ca II ratio as a function of electron
density ($n_e$) and temperature (T).  We restrict the temperature
range to 3.9 $<$ log T $<$ 4.1.  At lower temperatures a
\ion{Ca}{1}:\ion{Ca}{2} ratio above 0.1 would imply that the
\ion{Ca}{1}] line at $\lambda$6572 line would exceed the intensity of
the observed 6560~\AA\ feature (based on the \ion{Ca}{1}] excitation
rate from \citealt{SB01}).  At higher temperatures Ca would be ionized
to \ion{Ca}{3} and ions such as \ion{O}{2} and \ion{S}{2} are expected
to be strong.  None of these higher ionization states are observed
(see Figures~\ref{fig:spectra} and \ref{fig:nirspec}). The allowed
density range within these constraints is $10^{5.9}$\,cm$^{-3}$ at log
T = $4.1$ to $10^{6.7}$\,cm$^{-3}$ at log T = $3.8$.

Between day 95 and day 158, the 7300 \AA\/ flux drops by a factor of
4.6, while the Ca triplet flux drops by a factor of 20.  Homologous
expansion during that interval implies that the density drops by a
factor of 4.6.  The flux changes in both lines can be matched with a
factor of 4.6 drop in density and a factor of about 1.4 drop in
temperature.  That is a smaller change than the factor of 2.8 from
adiabatic expansion alone, indicating that radioactive heating partly
compensates for adiabatic cooling.

The [\ion{O}{1}] and [\ion{Ca}{2}] lines have similar excitation
potentials and critical densities, thus their intensity ratio depends
relatively weakly on temperature and density.  In the temperature
range derived above, most of the oxygen is neutral and most of the
calcium is \ion{Ca}{2}, so that the intensity ratio of the lines
reflects the Ca to O abundance ratio.  Since the [\ion{O}{1}] to
[\ion{Ca}{2}] ratio is nearly constant at $\approx 0.1$ during our
observations, but the emissivity of [\ion{Ca}{2}] is much larger than
that of [\ion{O}{1}] according to CHIANTI version 8
\citep{DelZanna15}, the O:Ca abundance ratio (by number) is
39$\pm$24. The substantial range is due to the range of temperature
and density derived above.

The relative excitation rates of [\ion{S}{1}] and [\ion{Ca}{2}] vary
less with density and temperature than those of [\ion{O}{1}] and
[\ion{Ca}{2}], but the uncertainty in the flux calibration of the NIR
spectrum and the necessity to interpolate the [\ion{Ca}{2}] fluxes
between nearby observations dominate the measurement uncertainty.  We
also assume that 1/5 of the 1.089 $\mu$m emission comes from He I to
match the 1.089 to 1.131 $\mu$m intensity ratio.  We use CHIANTI
version 8 to predict the \ion{S}{1} emissivities.  Notably, Table 3 of
\citet{Tayal04} mislabels the collision strengths from levels 1 and 3
to levels 4 and 5, and we have corrected the corresponding data file.
We find an S:Ca abundance ratio of 29$\pm$7.

The [\ion{Si}{1}] lines suffer from a still higher calibration
uncertainty and greater difficulties in removing night sky lines.
Using the excitation cross section of \citet{Pindzola77} and an
Einstein A value of 0.0037, we find that the [\ion{Si}{1}] emissivity
is 5 to 10 times smaller than the [\ion{S}{1}] emissivity over the
range of density and temperature that we infer.  The [\ion{Si}{1}]
flux is 1/3 that of the [\ion{S}{1}] lines, suggesting the Si
abundance is 1.7 to 3.3 times as large as that of S.  However, since
the [Si I] excitation rate did not include resonances or cascades, it
may be an underestimate.

At a distance of 30.4 Mpc, the [\ion{Ca}{2}] luminosity is
$3.5 \times 10^{40}~\rm erg~s^{-1}$.  With emissivities of
$2.0 \pm 0.6 \times 10^{-13}~\rm erg~s^{-1}$ per atom, that implies a
calcium mass of 0.006$\pm$0.002 M$_{\odot}$.  The masses of oxygen,
silicon, and sulfur are about $0.09_{-0.08}^{+0.11}$, $0.31 \pm 0.2$,
and 0.14$\pm 0.08$ M$_{\odot}$, respectively.

If we assume that the nebular emission arises in a shell given by a
velocity of 3400 $\rm km~s^{-1}$, or R=$2.8 \times 10^{15}~\rm cm$ at
day 95, and that the thickness of the shell is R/8 because of a steep
($n \propto R^{-9}$) density distribution, the density of \ion{Ca}{2}
ions is $5 \times 10^6~\rm cm^{-3}$.  As this lies within the electron
density range derived above, it is at least plausible that calcium
provides most of the free electrons in the nebular shell.

There are many caveats to our estimates. Absorption may obscure
interior emission \citep{Taubenberger09,Milisavljevic10}, the ejecta
may be clumpy \citep{Spyromilio94,Milisavljevic12}, and our assumption
that all lines are formed in the same volume may be incorrect. Indeed,
indications that some degree of asymmetry in the ejecta not accounted
for in our estimates can be seen in the clear asymmetry in the line
profiles (Figure~\ref{fig:lineprofiles}). Furthermore, there is
noticeable discrepancy between the masses we obtain from fitting
the light curve (section~\ref{sec:lc}) and from the spectra.  Our
spectral analysis may not be accounting for all mass because much of
the mass is at undetectable low densities and/or temperatures. This is
especially true for H and/or He material that may be neutral and at
temperatures that are too low for any significant excitation.
Improved chemical abundances and mass estimates may be possible with a
more sophisticated model that includes radiative transfer and
additional assumptions about ejecta kinematics.

\subsection{Metallicity and star formation rate}
\label{sec:metallicity}

Unlike many Ca-rich transients that are discovered in isolated
locations, iPTF15eqv is within projected proximity of a fairly rich
environment of hot gas and stars (Figure~\ref{fig:fchart}). Using our
optical spectra we estimated the metallicity of the region surrounding
the explosion site of iPTF15eqv using the Markov Chain Monte Carlo
method of \cite{Sanders12} to model the emission line spectra and
measure the fluxes of the strong emission lines. Following this
approach we estimate the oxygen abundance as
$\log(\rm{O} / \rm{H})+12=8.65\pm0.04$ and $8.58\pm0.02$ using the N2
and O3N2 diagnostics, respectively, of \cite{PP04}, and as
$\log(\rm{O} / \rm{H})+12=8.82\pm0.07$ using the diagnostic of
\cite{KD02} (KD02).  These metallicity estimates are consistent given
the characteristic offsets between the diagnostics \citep{KE08} and
uncertainties, and correspond to a metallicity of
$\sim0.8~\rm{Z}_\odot$. The quoted uncertainty is not inclusive of
systematic effects such as the calibration variance in the strong line
metallicity diagnostics ($\sim0.1$~dex, \citealt{KE08}) or local
metallicity variation in the host galaxy (as much as $0.3$~dex,
\citealt{Sanders12} and \citealt{Sanders12M31}).

NGC\,3430 is in the imaging footprint of the SDSS, which allowed us to
make use of the SDSS-calibrated Luminosity-Metallicity-Color relation
(LZC) of \cite{Sanders13} to infer an independent estimate of galaxy
metallicity from optical luminosity and color.  Using the extinction
corrected absolute magnitudes of the host galaxy in the SDSS DR12
cModel magnitudes \citep{Schlafly11,SDSSDR12}, $\rm{M}_g=-19.439$~mag
and $\rm{M}_r=-20.042$~mag, the LZC photometric metallicity for
NGC~3430 is $\log(\rm{O} / \rm{H})+12=8.68$ with a characteristic
scatter of $0.07$~dex in the \cite{PP04} N2 diagnostic calibration of
the LZC relation.

We also estimated the star formation rate (SFR) density of the
explosion site using the [\ion{O}{2}] and H$\alpha$ line fluxes
measured from our optical spectra.  We measure an [\ion{O}{2}]
$\lambda$3727 brightness of
$9.7\times10^{40}~\rm{erg}~\rm{s}^{-1}~\rm{arcsec}^2$ and an H$\alpha$
luminosity of $1.3\times10^{41}~\rm{erg}~\rm{s}^{-1}~\rm{arcsec}^2$,
with an estimated uncertainty in our spectrophotometric flux
calibration of $20\%$.  Using the [\ion{O}{2}] SFR diagnostic of
\cite{Kewley04}, assuming a distance scale of
$0.102~\rm{kpc}~\rm{arcsec}^{-1}$, and assuming the KD02 metallicity
quoted above, we find an explosion site SFR density of
$(60\pm10)\times10^{-3}~\rm{M}_\odot~\rm{yr}^{-1}~\rm{kpc}^{-2}$.  Using
the H$\alpha$ diagnostic of \cite{Moustakas06}, we find a SFR density
of $(80\pm20)\times10^{-3}~\rm{M}_\odot~\rm{yr}^{-1}~\rm{kpc}^{-2}$.
Given the uncertainty in the flux calibration, the two estimates are
similar.  Both approaches indicate a vigorous SFR consistent with the
morphological type of the galaxy.

\begin{figure}[tp]
\centering

\includegraphics[width=\linewidth]{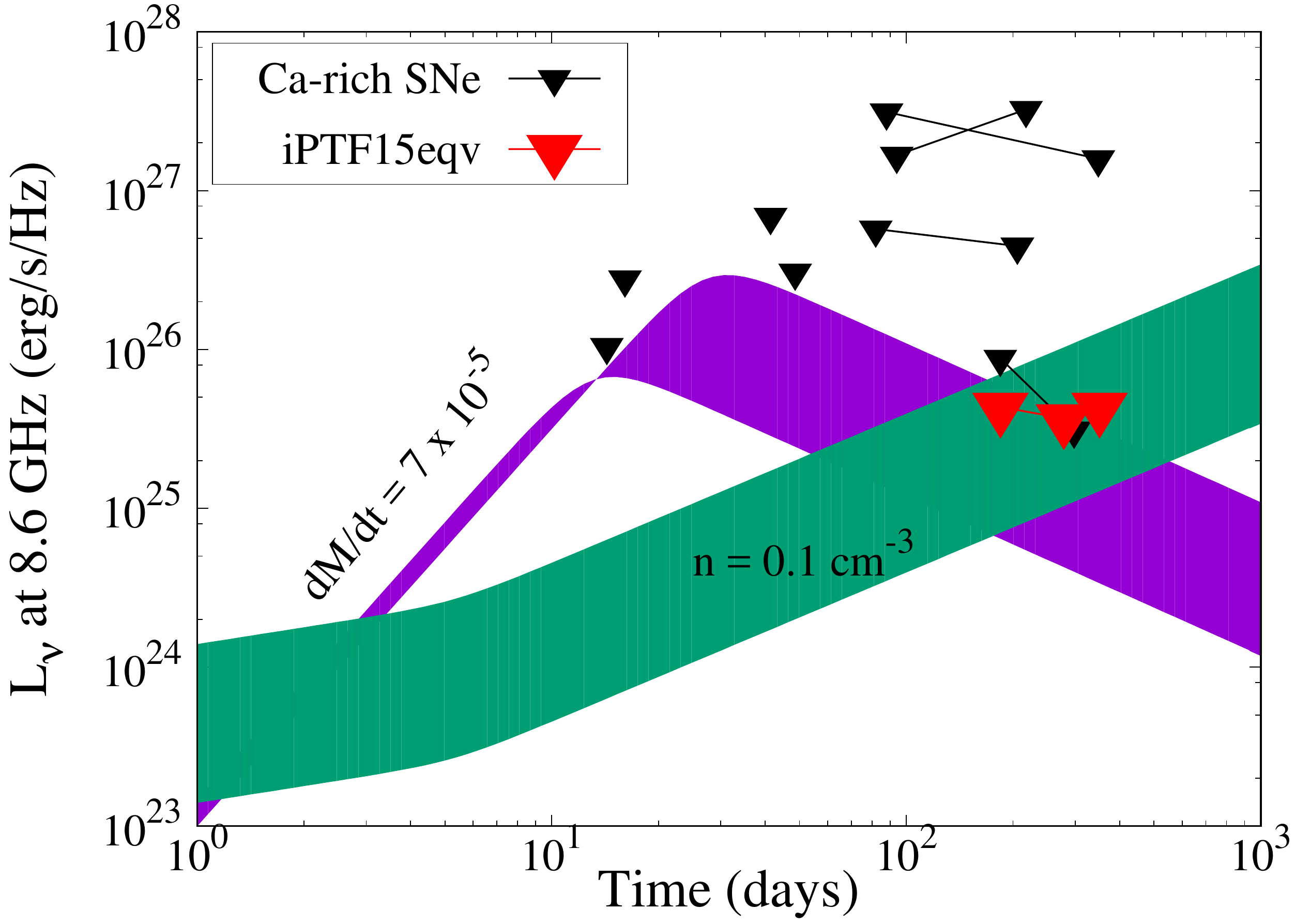}

\caption{Upper limits of radio emission at 8.6 GHz from iPTF15eqv
  (red) compared to those of other Ca-rich transients published in
  \citet{Chomiuk16} (black). The purple shaded region represents radio
  luminosity due to the wind-like medium with
  $\dot{M} = 7 \times 10^{-5}$ for the progenitor wind velocity of 100
  \kms. The green shaded region represents an ISM-like medium with
  density of 0.1 $\rm cm^{-3}$. The width of the shaded regions
  represents the spread due to the $0.01 \leq \epsilon_{B} \leq 0.1$
  and $\epsilon_{e} = 0.1$. We used $\rm E_{K} = 10^{51}$ erg and
  $\rm M_{ej} = 2 M_{\odot}$ (section~\ref{sec:lc}).}
\label{fig:radio}
\end{figure}

\subsection{Radio synchrotron limits}

The radio emission in SNe is due to synchrotron radiation. As the
forward shock wave moves through surrounding circumstellar and
interstellar material (CSM and ISM), a fraction of its kinetic energy
gets converted into motion of the shock-accelerated particles. The
relativistic electrons behind the shock are the main radiators which
gyrate in the shock-amplified magnetic field and radiate synchrotron
radiation \citep{Chevalier82}. The energy distribution of these
relativistic electrons is a power law,
$n_{e}(\gamma_{e}) \propto \gamma_{e}^{-p}$, with index $p$ and the
Lorentz factor $\gamma_{e}$ of individual electrons. This form of the
electron distribution is often evident from the synchrotron radio
spectrum of SNe.  Thus, we can utilize our deep radio observations of
iPTF15eqv to probe unique information about its immediate environment.

For a constant mass-loss rate and wind velocity, the density in the
stellar environment falls with the distance as the inverse squared
$\rho \propto r^{-2}$. In a typical massive star, this wind-density
profile may extend to several parsecs. A typical supernova shock wave
traveling at the speed of a few $\times 10^{4}$~\kms\ may take several
decades before crossing the progenitor wind.

We followed procedures outlined in \citet{Kamble14}, based on work
from \citet{Chevalier82}, \citet{Chevalier98}, and
\citet{Chevalier06}, to estimate the radio emission generated by a
shock wave expanding in such an environment.  The deep upper limits on
radio brightness evolution at 8.6 GHz constrain the progenitor wind
mass-loss rate to be
$\dot{M} \la 7 \times 10^{-5}\, \rm M_{\odot}\;yr^{-1}$ assuming a
constant wind velocity of $v_{w} = 100$ \kms. This is shown in
Figure~\ref{fig:radio} by the purple shaded region. We also show the
expected radio light curves for the ISM density profile. Our
observations constrain the ISM density to be $\rm n \la 0.1$
cm$^{-3}$. The width of the shaded regions represents the spread due
to the $0.01 \leq \epsilon_{B} \leq 0.1$ and $\epsilon_{e} = 0.1$,
where $\epsilon_{B}$ and $\epsilon_{e}$ are the fractions of kinetic
energy channeled into the post-shock magnetic fields and shock
accelerated relativistic electrons, respectively.

\begin{figure*}[tp]
\centering

\includegraphics[width=0.95\linewidth]{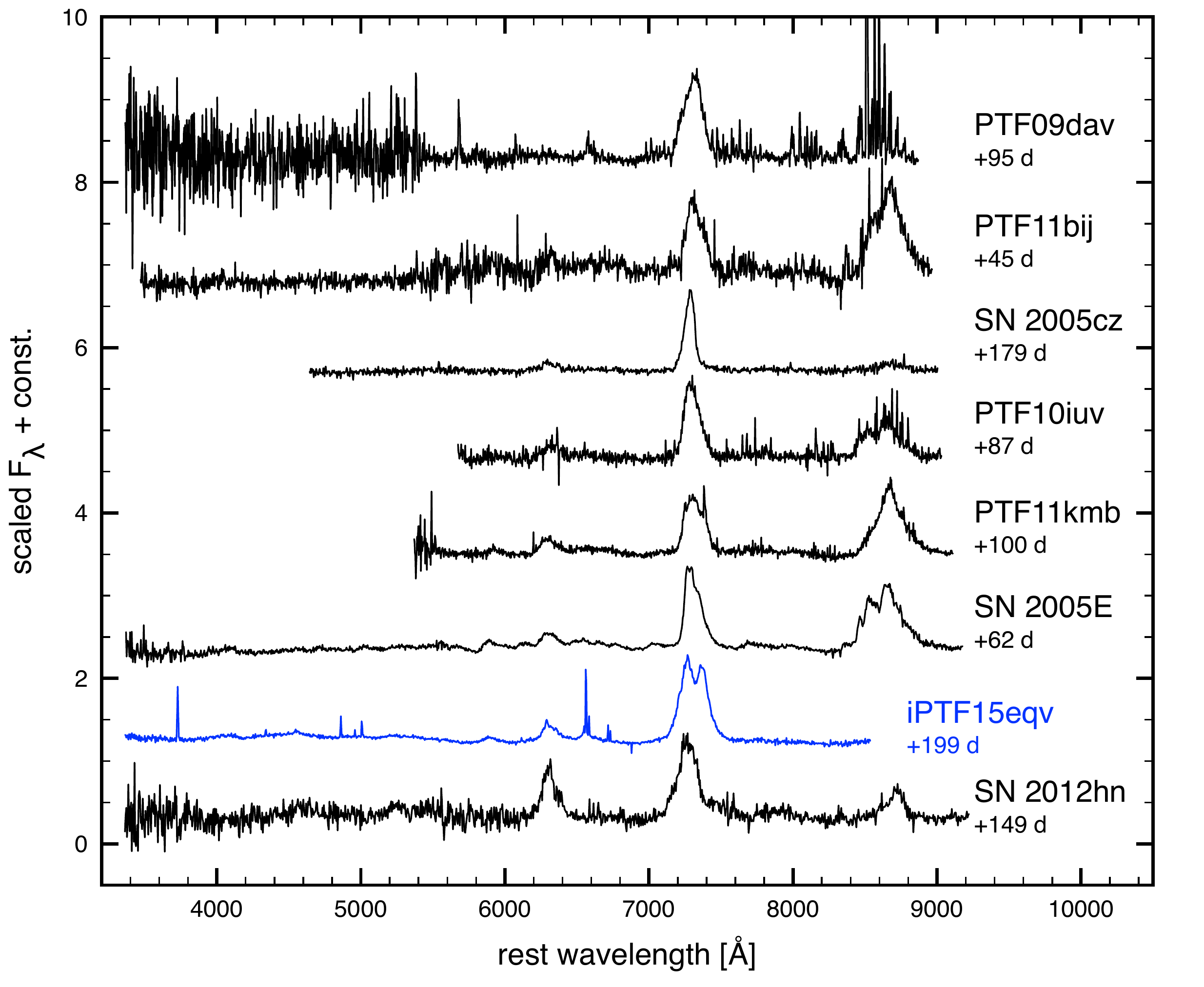}

\caption{Optical spectra of Ca-rich transients compared to that of
  iPTF15eqv. Archival data were originally published in
  \citet{Kawabata10}, \citet{Perets10}, and \citet{Kasliwal12} and
  have been obtained through WISEREP \citep{YG12} and the OSC
  \citep{Guillochon17}.}

\label{fig:ca-rich-compare}
\end{figure*}

\subsection{Tidal disruption}

One model of Ca-rich transients is that they are tidal disruptions of
low mass WDs by IMBHs
\citep{Rosswog08,Metzger12,MacLeod14,Sell15,Tanikawa17}. \citet{MacLeod16}
provide observable signatures of this process at optical
wavelengths. We find that our overall light curve evolution is not
consistent with extrapolations of their models that generally show
much flatter and nearly constant evolution in $R$- and $I$-band, and
we do not have photospheric epoch spectra to compare with their
synthetic spectra.

However, a unique signature of tidal detonations by IMBHs is that they
are accompanied by the fallback of substantial amounts of stellar
debris returning to the BH that can be accreted and emit at
X-ray wavelengths. Our CXO observations of iPTF15eqv can be used to
constrain the likelihood of this scenario.

The fallback rate can be several orders of magnitude above the
Eddington limit \citep{MacLeod16}, with the inner X-ray emitting
portions of the accretion structure being potentially obscured by
optically-thick material. While a jet can be produced during this
hyper-accretive phase, the absence of any detected afterglow suggests
that no jet was formed. The disk produced by the fallback will be able
to radiate at roughly its Eddington limit during this phase,
\begin{equation*}
L_{\rm Edd} \equiv G M_{\rm h} m_{\rm p} c / \sigma_{\rm T} \simeq
10^{41}~M_{\rm h,3}\,{\rm erg~s}^{-1},
\end{equation*}
where $M_{\rm h,3}$ is the BH mass in units of $10^3
\rm M_\odot$.
Using the peak accretion rate and time found in simulations
\citep{Guillochon13}, the timescale that the fallback rate will remain
above the Eddington limit is
\begin{equation*}
t_{\rm Edd} = 1~M_{\rm h, 3}^{-2/5} M_{\rm WD,0.6}^{1/5} R_{\rm
  WD,0.6}^{3/5}~\rm yr,
\end{equation*}
where $M_{\rm WD,0.6}$ and $R_{\rm WD,0.6}$ are the mass and radius of
an $0.6 \rm M_\odot$ WD. \citeauthor{MacLeod16} make the point that
viscous effects may act to spread the accretion of matter over
somewhat longer timescales, prolonging the time that the accretion
rate remains above Eddington up to a factor of $\sim 10$. After this
phase, the accretion luminosity should decay approximately as
$(t/t_{\rm peak})^{-5/3}$ \citep{Rees88}.

\citet{Sell15} observed SN\,2012hn 533 days after the explosion to
look for accretion emission during the decay phase. No emission was
detected, and from an X-ray upper limit on the luminosity of
$L_{\rm 0.5-8kev} < 4.4 \times 10^{38}\; {\rm erg\;s}^{-1}$ they
concluded that SN\,2012hn was unlikely to be a tidal detonation by an
IMBH unless the BH is at the low end of the IMBH mass
distribution. Our upper-bound limit on the luminosity
($<5.3\times 10^{38}\,\rm{erg\,s^{-1}}$) of iPTF15eqv is comparable to
that of SN\,2012hn, but since our observation of iPTF15eqv was closer
to the time of explosion than SN\,2012hn (114 -- 164 days versus 533
days) we can place even stricter constraints on the IMBH mass. Given
that we would expect the accretion disk to be radiating at close to
its Eddington limit for nearly a year and decline rapidly afterwards,
and that even the 10\% accretion efficiency expected in the low-hard
state \citep[as assumed in][]{Sell15} would have been detected, we can
restrict the BH mass to the point that only a stellar mass BH
($\lesssim 100 M_\odot$) could yield $t_{\rm Edd} < 164~{\rm days}$.

Our constraint on the BH mass falls into the regime of a ``micro''
tidal disruption event \citep{Metzger12,Perets16}, a scenario that
would only be likely in a dense stellar cluster where close encounters
between stars are common.  In these models a sufficiently massive WD
is tidally disrupted by its BH companion as the binary orbit shrinks
due to unstable mass transfer. The WD debris is sheared into an
accretion disk with an initial size that is comparable to that of the
initial binary at the time of Roche-lobe overflow.

\section{Discussion}
\label{sec:discussion}

iPTF15eqv exhibits a unique combination of properties that bridge
those observed in Ca-rich transients and SNe Ib/c. On the one hand,
the late-time spectra of iPTF15eqv are dominated by strong calcium
emission lines, which is a defining characteristic of Ca-rich
transients. We illustrate this similarity in
Figure~\ref{fig:ca-rich-compare}, where we show the day 199 spectrum
of iPTF15eqv and a sample of Ca-rich transients from the literature.
On the other hand, iPTF15eqv differs from other Ca-rich transients in
terms of its light curve evolution. iPTF15eqv was a slower evolving
and potentially much more luminous object than Ca-rich transients and
exhibited a light curve that resembles SNe Ib/c (Figure
~\ref{fig:ca-rich-lc}). Below we explore how to place iPTF15eqv within
the Ca-rich transient classification and seek to understand whether a
single progenitor system framework can unite their diverse properties.

\begin{figure}[tp]
\centering

\includegraphics[width=\linewidth]{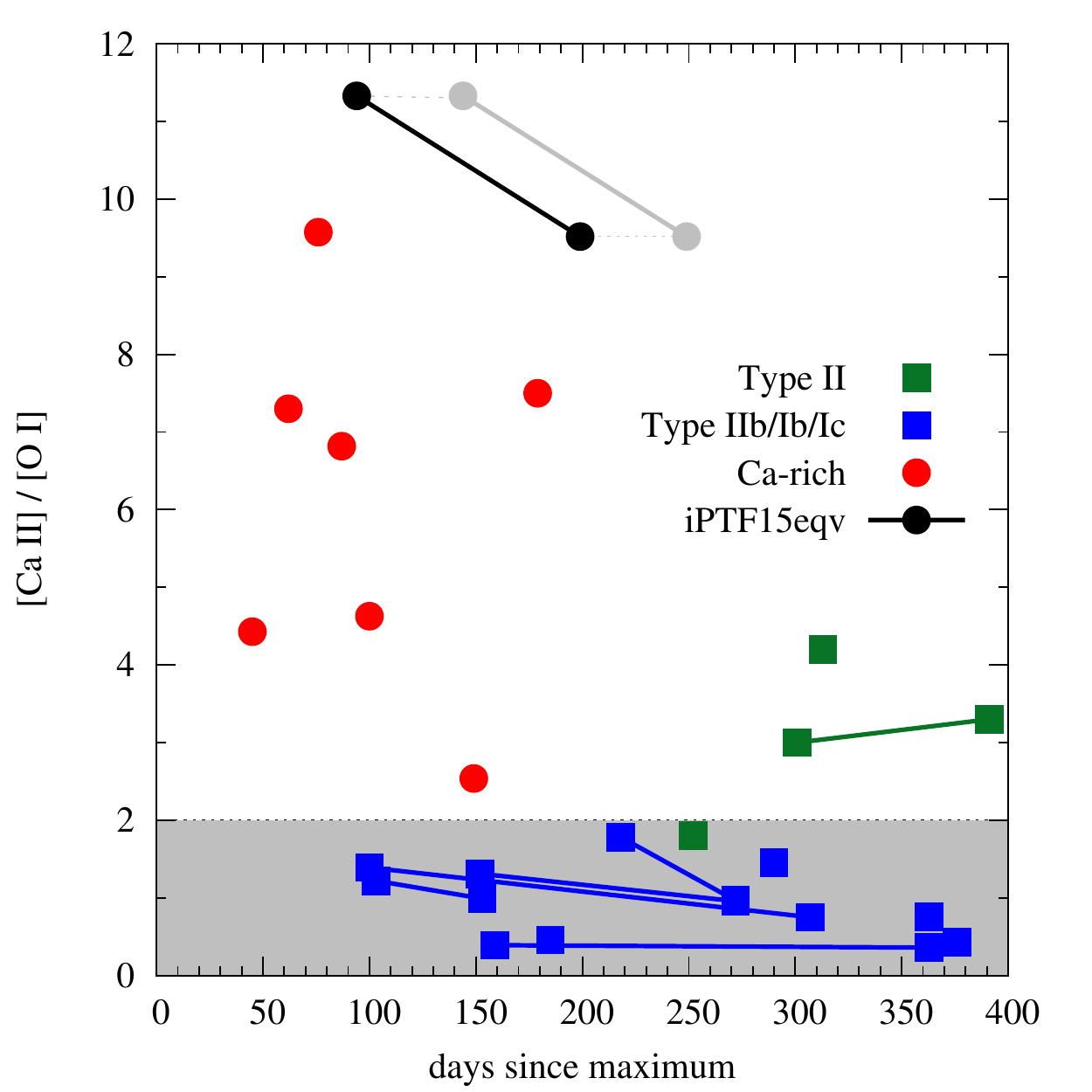}

\caption{Ratio of [\ion{Ca}{2}] $\lambda\lambda$7291, 7324 /
  [\ion{O}{1}] $\lambda\lambda$6300, 6364 for Type II, Type IIb/Ib/Ic,
  and Ca-rich transients. At all epochs where both lines are visible
  and the conditions can be reasonably assessed to be nebular, all
  Ca-rich transients have [\ion{Ca}{2}]/[\ion{O}{1}] $> 2$, and all
  Type Ib/c are $<2$. Solid lines connect different epochs of the same
  object. The gray silhouette of iPTF15eqv connected by dashed lines
  reflects the uncertainty in the explosion date. Refer to Table~4 for
  data used to create this plot.}

\label{fig:CaIIvOI}
\end{figure}

\subsection{[\ion{Ca}{2}]/[\ion{O}{1}]}

For Ca-rich transients having optical spectra publicly accessible
through the Weizmann Interactive Supernova Data Repository (WISEREP;
\citealt{YG12}) and the Open Supernova Catalog (OSC;
\citealt{Guillochon17}), we measured the [\ion{Ca}{2}]/[\ion{O}{1}]
relative flux ratio (for cases where both lines could be observed) to
determine the range of ratios encountered thus far in the
literature. Multiple epochs were measured when possible. The results
are listed in Table~4 and plotted in Figure~\ref{fig:CaIIvOI}. For
comparison, we also measured this ratio for a number of H-rich (Type
II) and H-poor (Type IIb/Ib/Ic) SNe. At all epochs where both lines
are visible and the conditions can be reasonably assessed to be
nebular, all Ca-rich transients have [\ion{Ca}{2}]/[\ion{O}{1}] $> 2$
and all Type Ib/c are $<2$. iPTF15eqv exhibits the strongest
[\ion{Ca}{2}]/[\ion{O}{1}] emission ratio ($\approx 10$) in our
sample. SNe II typically have higher [\ion{Ca}{2}]/[\ion{O}{1}] ratios
between $\approx 2-4$. However, the phase at which this ratio is
observed in Type II supernovae ($ t > 200$ d) is much later than
Ca-rich transients ($50 < t < 150$ d).

Notably, the ratio of [\ion{Ca}{2}]/[\ion{O}{1}] for many Type Ib/c
supernovae can decrease with time. For example, a 30\% change in the
ratio was noted in SN\,2004gq between day 258 and 381
\citep{Kuncarayakti15}. For this reason, \citet{Valenti14} raised the
issue that the high [\ion{Ca}{2}]/[\ion{O}{1}] ratio seen in some
Ca-rich transients at relatively early phases may in part be a
phase-dependent effect rather than reflecting an intrinsic abundance
pattern.

In the case of iPTF15eqv, the [\ion{Ca}{2}]/[\ion{O}{1}] ratio
exhibited modest change between the the epochs where both features
were well observed (i.e., days 95-199; Figure~\ref{fig:spectra}).
This lack of change is consistent with observations of SN Ib/c at very
evolved epochs where change is routinely negligible
\citep{Elmhamdi04}, and in line with the fact that the critical
densities for [\ion{O}{1}] and [\ion{Ca}{2}] are comparable. Although
we have not explored epochs $t > 300$ d, a decrease in
[\ion{Ca}{2}]/[\ion{O}{1}] ratio alone appears to be an unlikely
explanation for why Ca-rich transients have strong calcium line
emission.

\begin{figure*}[tp]
\centering

\includegraphics[width=0.49\linewidth]{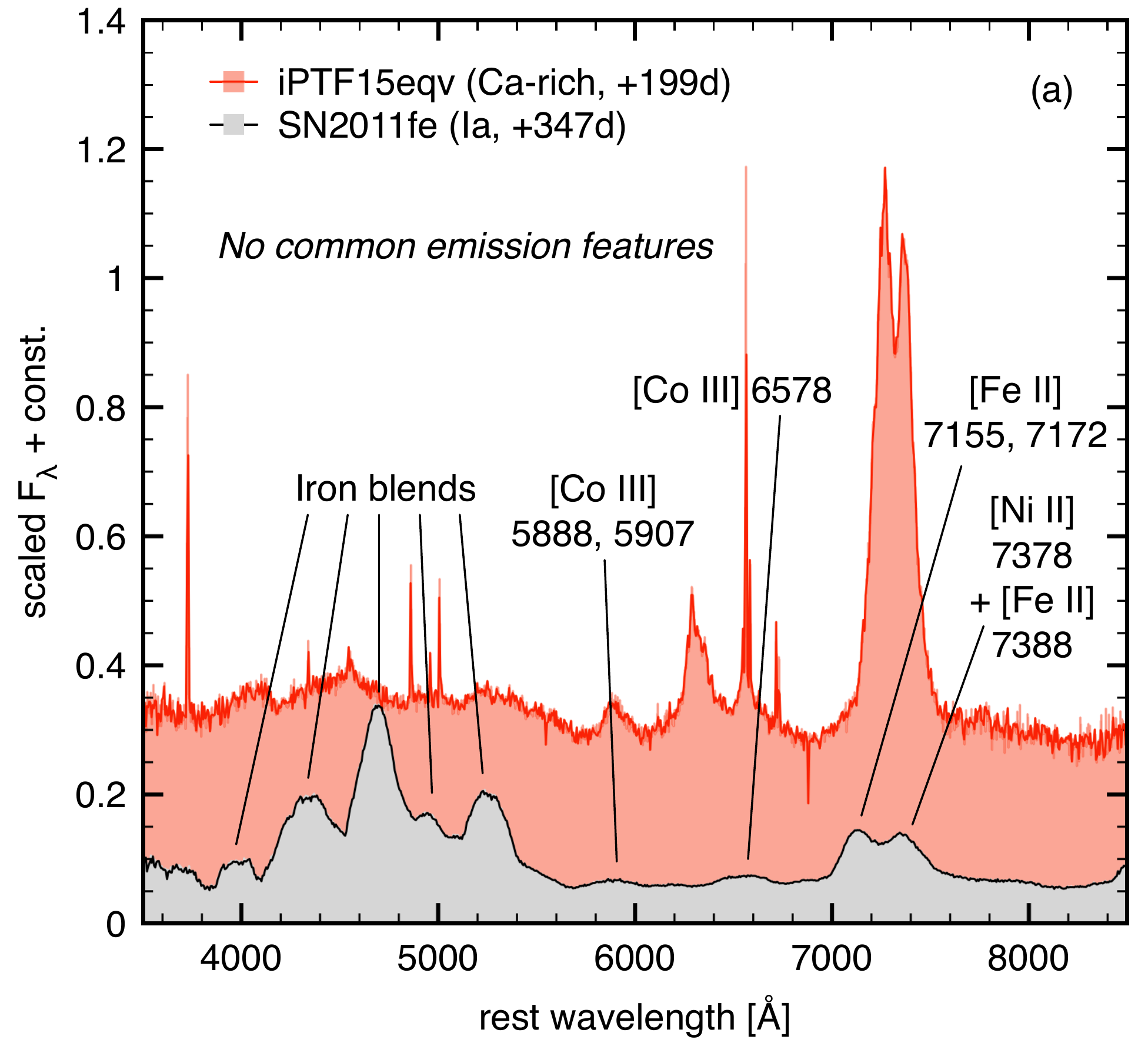}
\includegraphics[width=0.49\linewidth]{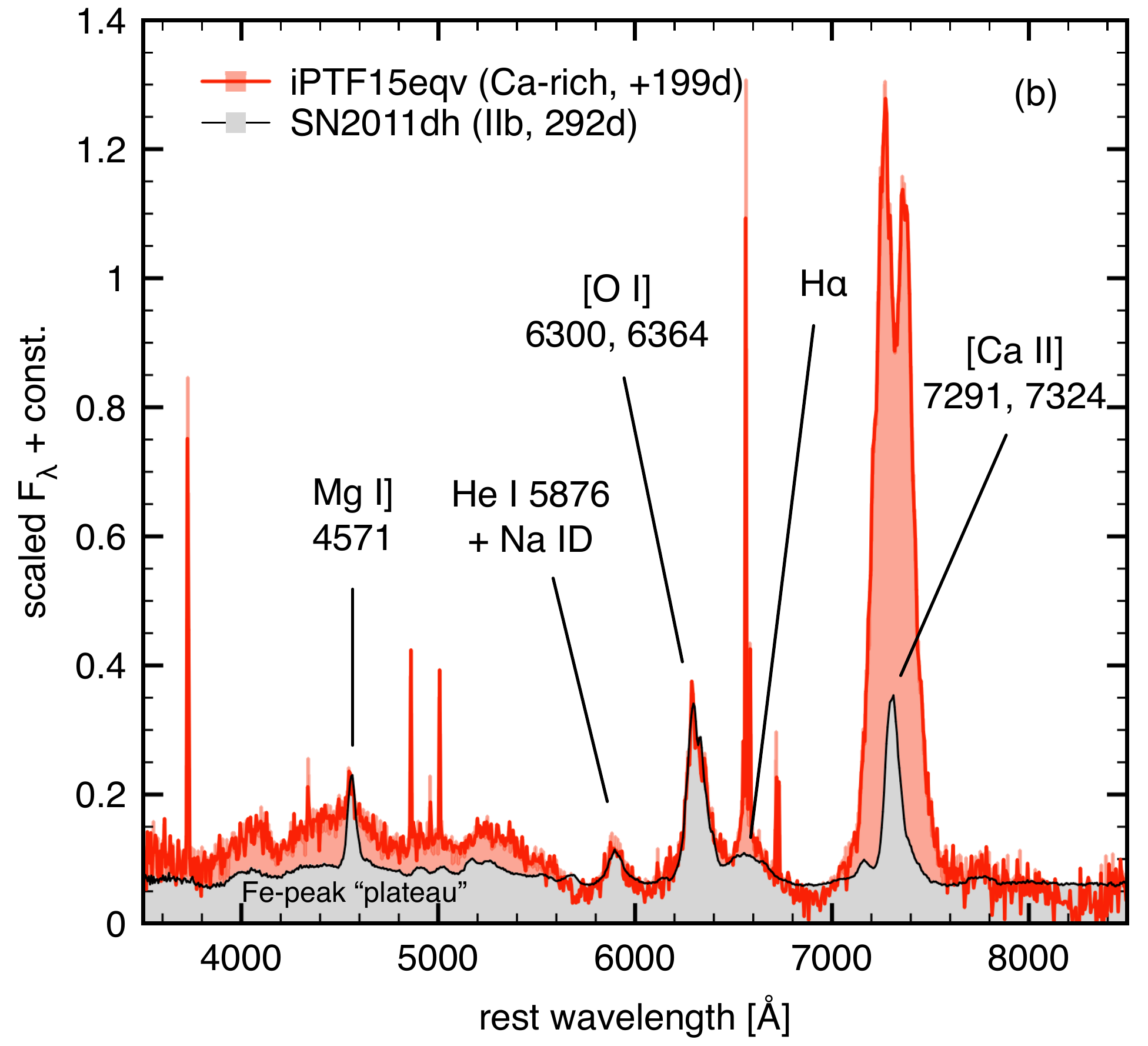}\\
\includegraphics[width=0.49\linewidth]{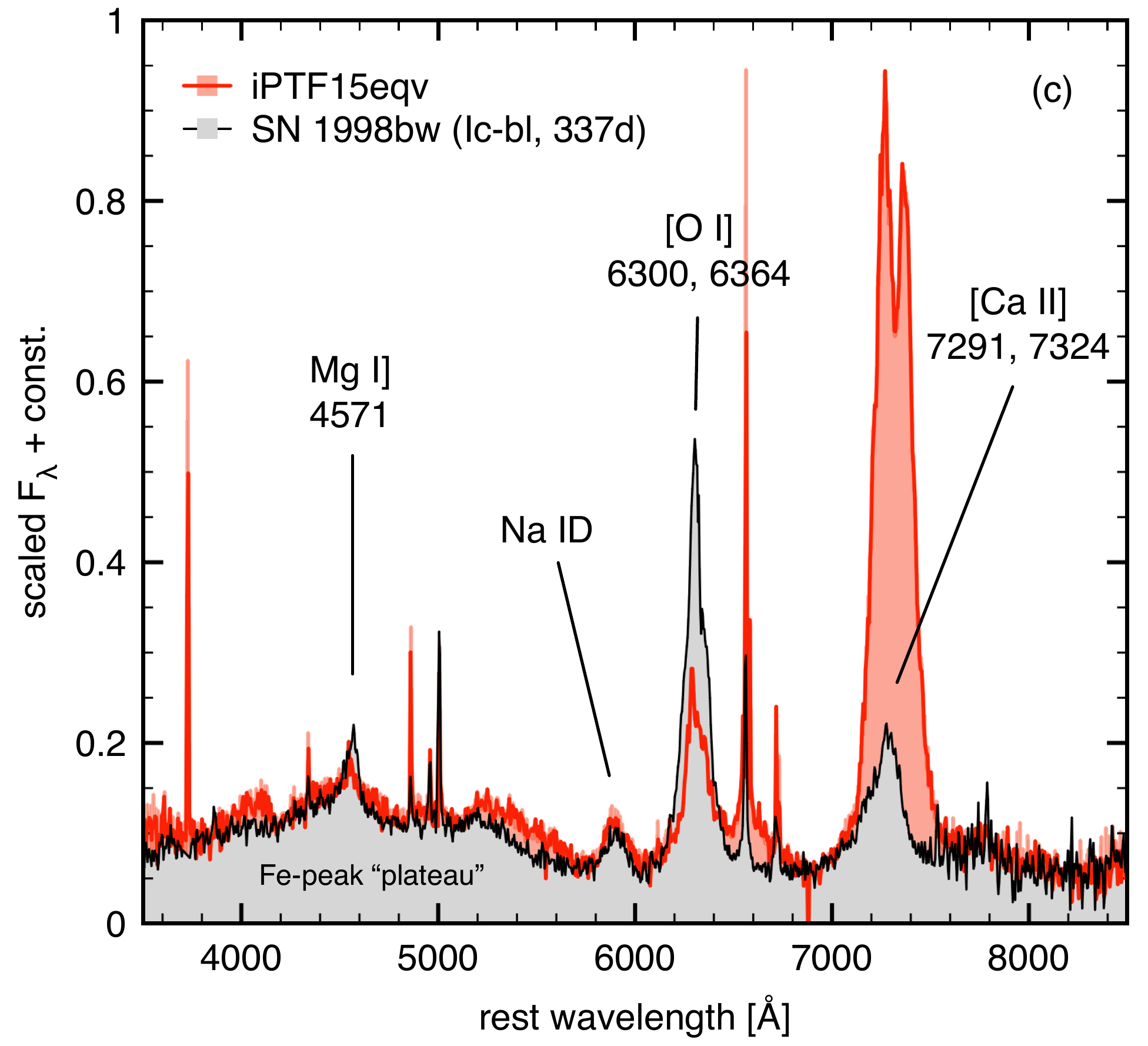}
\includegraphics[width=0.49\linewidth]{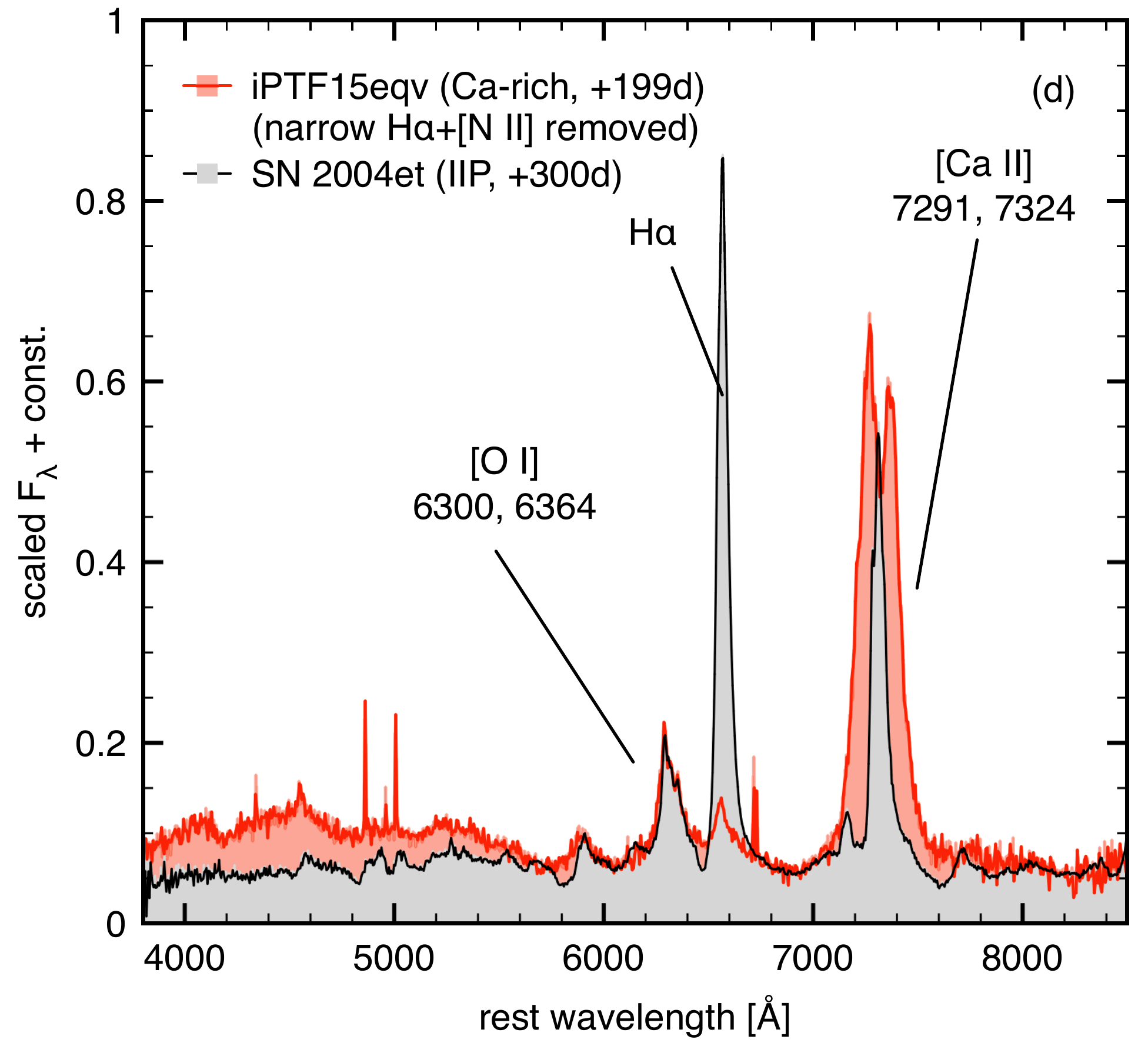}

\caption{Spectral fingerprinting iPTF15eqv. Spectra have been
  arbitrarily scaled to compare and contrast the relative strengths of
  specific emission features between nebular epoch spectra of
  iPTF15eqv and other supernovae (both thermonuclear and core
  collapse) The spectral features of iPTF15eqv best match those of a
  core-collapse explosion and have little similarity to those observed
  in thermonuclear WD explosions. Data are originally published in
  \citet{Mazzali15}, \citet{Ergon15}, \citet{Patat01}, and
  \citet{Sahu06}.}

\label{fig:haha}
\end{figure*}

\begin{figure*}[tp]
\centering

\includegraphics[width=0.49\linewidth]{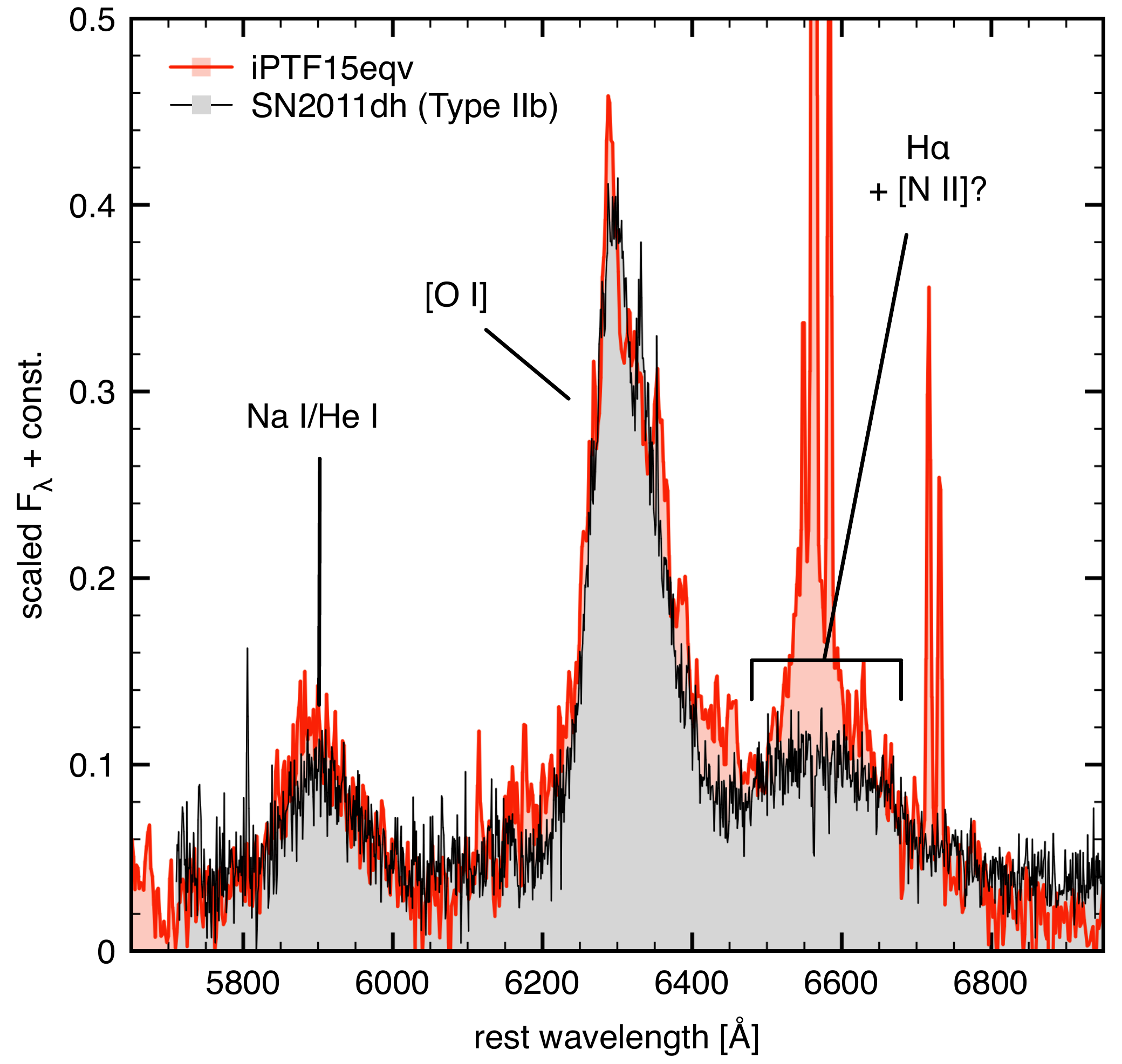}
\includegraphics[width=0.49\linewidth]{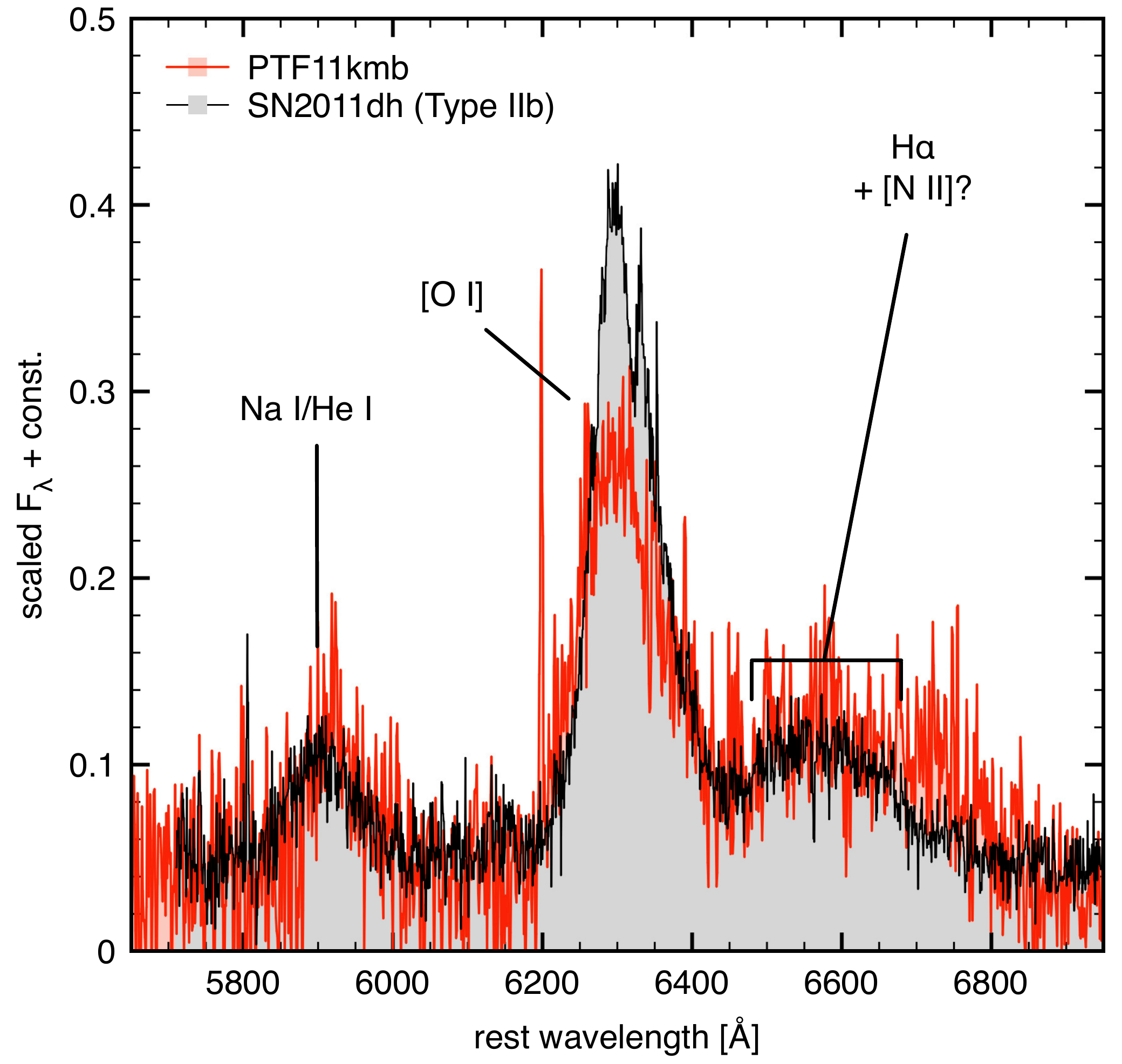}

\caption{Optical spectra of Ca-rich transients iPTF15eqv and PTF11kmb
  enlarged in the region of 5800-6800~\AA\ and scaled
  arbitrarily. Broad emission centered around 6560~\AA\ is similar in
  distribution and strength to that observed in the Type IIb
  SN\,2011dh \citep{Ergon15}, where it is attributable to H$\alpha$
  and [\ion{N}{2}] $\lambda\lambda$6548, 6583. Emission from the
  \ion{Ca}{1}] $\lambda$6572 line may also contribute to this feature
  in Ca-rich transients.}

\label{fig:sn2011dh}
\end{figure*}

\subsection{Spectral fingerprinting}
\label{sec:late-time}

Optical and NIR spectroscopy obtained many months to years after
maximum light is a powerful probe of the kinematic and chemical
properties of the ejecta closest to the explosion center. It can be
used to discriminate between Type Ia versus core-collapse
classifications, and even between sub-classifications
\citep{Milisavljevic12,Parrent+16}, and thus provides a unique means
of spectrally fingerprinting a supernova's progenitor system.

Few Ca-rich transients have had high-quality, multi-epoch data at late
times.  The close proximity and relatively slow decline rate of
iPTF15eqv enabled spectroscopic observations of high signal-to-noise
ratio at epochs $>$ 200 days after explosion, which is among the
latest epochs ever observed for a Ca-rich transient. The high quality
of the spectra reveals faint features that would be missed in lower
signal-to-noise data.

We compared the day 199 spectrum of iPTF15eqv to prototypical Type I
and II SNe (Figure~\ref{fig:haha}). Spectra were arbitrarily scaled to
compare and contrast the relative strengths of specific emission
features between nebular epoch spectra of iPTF15eqv and other
supernovae. Clearly, iPTF15eqv and the normal, well-observed Type Ia
SN\,2011fe have widely different late-time emissions. Type Ia
late-time spectra are dominated by a complicated blend of forbidden
and permitted iron lines between 4000 and 5500~\AA, many of which are
associated with Fe-peak elements. Peaks centered around 7000-7600~\AA\
associated with iron and nickel are also regularly observed. None of
these features are conspicuous in iPTF15eqv.

In sharp contrast, iPTF15eqv shares many spectral signatures of
core-collapse SNe. The late-time spectrum of the Type IIb SN\,2011dh
\citep{Ergon15}, exhibits [\ion{O}{1}], [\ion{Ca}{2}], a blend of
\ion{He}{1}+\ion{Na}{1}, a ``plateau'' of emission spanning
3800-5600~\AA\ associated with Fe-peak elements, and a broad H$\alpha$
emission feature redward of [\ion{O}{1}] (that may have some
contribution from [\ion{N}{2}]), which overlap with iPTF15eqv. The
noticeable difference is in the strength of the Fe-peak plateau and
the [\ion{Ca}{2}]/[\ion{O}{1}] ratio, as well as the \ion{Mg}{1}]
emission which is prominent in many SNe~IIb but not so in iPTF15eqv.

A surprising similarity is also found between iPTF15eqv and
SN\,1998bw, which is a broad-lined Type Ic that was associated with
the long-duration $\gamma$-ray burst GRB 980425
\citep{Galama98,Patat01}. In this case, the relative strength of the
Fe-peak plateau is comparable between the two, but the
[\ion{Ca}{2}]/[\ion{O}{1}] is much smaller. Considerable overlap
between the late-time spectra of the Type IIP SN\,2004et
\citep{Sahu06} and iPTF15eqv is also evident. However, the H$\alpha$
line dominates the spectrum of SN\,2004et ([Ca II]/H$\alpha$ = 0.6),
and is far stronger relative to that observed in iPTF15eqv ([Ca
II]/H$\alpha$ = 30; but see also section \ref{sec:hydrogen} for
discussion of other emission lines that may contribute to this
wavelength region).

Notably, the [\ion{Ca}{2}] 7291, 7324 emission line profile of
iPTF15eqv is unlike any other Ca-rich transient (Figures
\ref{fig:lineprofiles} and \ref{fig:ca-rich-compare}), and rare among
SNe Ib/c. To our knowledge, the only instance is in late-time spectra
of the Type Ib SN\,2007uy, though in that case the velocity span was
much smaller ($\la 5000$ \kms;
\citealt{Milisavljevic10,Roy13,Modjaz14}). Double-peaked profiles have
been observed in the [\ion{O}{1}] $\lambda\lambda$6300, 6364 line
profiles of SNe Ib/c and interpreted to represent ejecta asymmetry
\citep{Modjaz08,Maeda08,Taubenberger09}, possibly in the form of an
O-rich torus \citep{Maeda07}, that may have been shaped by a jetted
explosion \citep{Maeda02,Mazzali05}. However, many of these cases are
actually due to the doublet nature of [\ion{O}{1}] in combination with
significant density enhancements associated with ejecta clumping
\citep{Milisavljevic10}.

\subsection{Hydrogen in Ca-rich transients}
\label{sec:hydrogen}

The strength of the 6560~\AA\ feature relative to [\ion{O}{1}]
observed in iPTF15eqv is broadly similar to that observed in
SN\,2011dh and other Type IIb supernovae.  We identify this feature to
be primarily H$\alpha$ in iPTF15eqv, however, we have noted that
\ion{Ca}{1}] 6572 and [\ion{N}{2}] 6548, 6583 may also contribute
emission in this region. In Figure~\ref{fig:sn2011dh} we show the day
199 spectrum of iPTF15eqv and a day 292 spectrum of SN\,2011dh. The
emission in iPTF15eqv is more narrowly peaked, which reflects
differences in how the presumably H-rich ejecta are distributed in the
two objects. A parabolic distribution is anticipated from a spherical
geometry. A flat-topped distribution as observed in SN\,2011dh is
reflective of a shell-like geometry.

iPTF15eqv is not unique in potentially exhibiting H$\alpha$ emission
among Ca-rich transients. The subluminous and peculiar PTF09dav was
reported to exhibit H$\alpha$ with a velocity width of $\sim 1000$
\kms\ in nebular spectra presented in \citet{Kasliwal12}. The
H$\alpha$ emission was interpreted as circumburst material being
photoionized from interaction between the explosion and the progenitor
star's wind. The \citet{Kasliwal12} scenario for PTF09dav is different
from what we propose for iPTF15eqv, where H$\alpha$ exhibits higher
velocities that are more easily understood as being associated with
explosive ejecta. The possibility of a contribution from \ion{Ca}{1}]
6572 was not discussed within the context of PTF09dav. In
Figure~\ref{fig:sn2011dh}, we also show the Ca-rich transient PTF11kmb
enlarged around the same region.  Intriguingly, PTF11kmb exhibits
excess emission redward of [\ion{O}{1}] that resembles the H$\alpha$
distribution of SN\,2011dh.

\begin{figure}[tp]
\centering

\includegraphics[width=\linewidth]{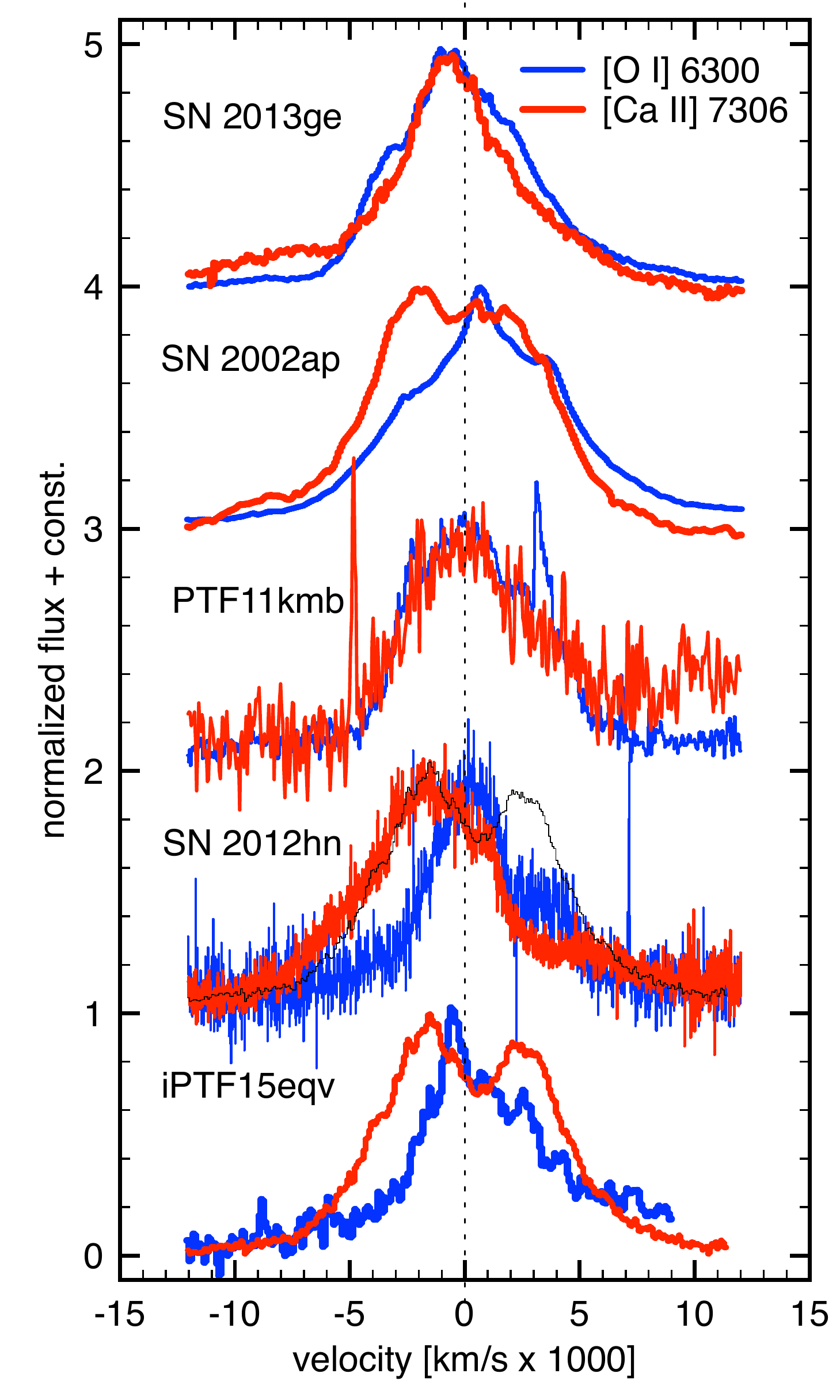}

\caption{Emission line profiles of [\ion{O}{1}] 6300, 6364 and
  [\ion{Ca}{2}] 7291, 7324 for SNe Ib/c and Ca-rich
  transients. Velocities are with respect to 6300~\AA\ and 7306~\AA,
  respectively. The [\ion{Ca}{2}] profile does not always overlap with
  [\ion{O}{1}], which makes it a poor tracer of the systemic velocity
  of bulk ejecta. Data were obtained from WISEREP and the OSC and were
  originally published in the following papers: SN\,2013ge
  \citep{Drout16}, SN\,2002ap \citep{Foley03}, PTF11kmb
  \citep{Kasliwal12}, SN\,2012hn \citep{Valenti14}, iPTF15eqv (this
  paper).}

\label{fig:shift}
\end{figure}

\subsection{High velocity kicks?}

To explain how Ca-rich transients explode in isolated regions devoid
of massive star populations, \citet{Lyman14} proposed long-lived
progenitor systems that have been ``kicked'' to high
velocities. Similarly, \citet{Foley15} argued that the progenitors of
Ca-rich transients are double WD systems that have been ejected from
their host galaxies after interacting with a super-massive
BH. \citet{Foley15} analyze the nebular spectra of 13 Ca-rich
transients to look for evidence in support of this hypothesis in the
[\ion{Ca}{2}] lines and find a correlation between projected distance
from the host galaxy and bulk Doppler shift velocity. Specifically,
SNe with small projected offsets have large line-of-sight velocity
shifts as determined by nebular lines, while those with large
projected offsets have no significant velocity shifts.

We considered the implications of \citet{Foley15} on the unusual
double-peaked [\ion{Ca}{2}] profile of iPTF15eqv and found an
important caveat in this interpretation. Although the [\ion{Ca}{2}]
line profile exhibits clear asymmetry potentially associated with a
bulk Doppler shift, the profiles of all other ions of iPTF15eqv are
centered near zero rest velocity (Figure~\ref{fig:lineprofiles}). This
characteristic of iPTF15eqv is not unique, in that the [\ion{Ca}{2}]
emission line profile of SN explosions at nebular epochs can often
have an apparent blueshift attributable to internal absorption of
inner ejecta \citep{Taubenberger09,Milisavljevic10}. If the ejecta are
not completely optically thin, then emission from the rear of the
expanding debris cloud is preferentially blocked. This effect, which
can affect some ions more than others, gives the overall impression of
being blueshifted.

We illustrate this phenomenon of apparent blueshifts in
Figure~\ref{fig:shift}, where we show the emission line profiles of
[\ion{Ca}{2}] and [\ion{O}{1}] for two Type Ib/c supernovae and two
Ca-rich events, in addition to that of iPTF15eqv.  [\ion{Ca}{2}]
emission has a large range distributions that sometimes reflects the
same distribution as [\ion{O}{1}] (SN\,2013ge and PTF11kmb), but other
times can show an apparent bulk blueshift (SN\,2002ap and
SN\,2012hn). \citet{Foley15} interpret the [\ion{Ca}{2}] profile of
SN\,2012hn to be associated with a systemic velocity of $-1730 \pm 70$
\kms. However, this view is not easily reconciled with the emission
line profile of [\ion{O}{1}] that is distributed about zero
velocity. Thus, we do not find strong evidence supporting a high
systemic bulk velocity in iPTF15eqv and conclude that other Ca-rich
transients may be similarly affected by optical depth effects.

Interestingly, the blueshifted [\ion{Ca}{2}] emission of iPTF15eqv and
SN\,2012hn trace similar distributions. SN\,2012hn, however, does not
have a prominent redshifted emission peak like that observed in
iPTF15eqv. This could be consistent with a shared torus geometry, if
in the case of SN\,2012hn its rear hemisphere has been obscured. This
speculative scenario suggests that an observer's line of sight may be
an important consideration in interpreting spectra of Ca-rich
transients.

\subsection{Metallicities of Ca-rich transient host galaxies}

Although indirect lines of evidence from host galaxy samples and
explosion site locations point toward low mass and low metallicity
progenitor stars \citep{Yuan13}, no systematic observational survey of
the host galaxy metallicities or explosion site star formation rates
of Ca-rich SNe has been conducted.

We examined 5 Ca-rich transients (SN~2000ds, 2001co, 2003dg, 2005E,
and PTF11bij) that have host galaxies in the SDSS imaging footprint
and for which we could apply the LZC-relation of \cite{Sanders13} to
estimate galaxy metallicity from optical luminosity and color.  The
foreground-extinction-corrected absolute SDSS magnitudes for each host 
galaxy and the host galaxy global metallicities derived from the
LZC relation on the N2 scale of \cite{PP04} are displayed in
Table~5. 

Host galaxies for four of the five Ca-rich transients (SN~2000ds,
2001co, 2003dg, and PTF11bij) have $\log(\rm{O} / \rm{H})+12\sim8.7$,
which is consistent with the metallicity reported here for
iPTF15eqv. SN~2005E is the sole outlier with a substantially lower
global metallicity ($\log(\rm{O} / \rm{H})+12=8.4$). Thus, we conclude
that Ca-rich transients often occur in metal-rich galaxies.  Notably,
this finding is not at odds with a low-mass star progenitor
scenario. The galaxy mass-normalized rate of Type~Ia supernovae is
actually highest, marginally, in Type~Sc host galaxies among all major
morphological types (\citealt{Li11Hosts}; see also \citealt{Lyman13}).

\begin{figure}

\includegraphics[width=\linewidth]{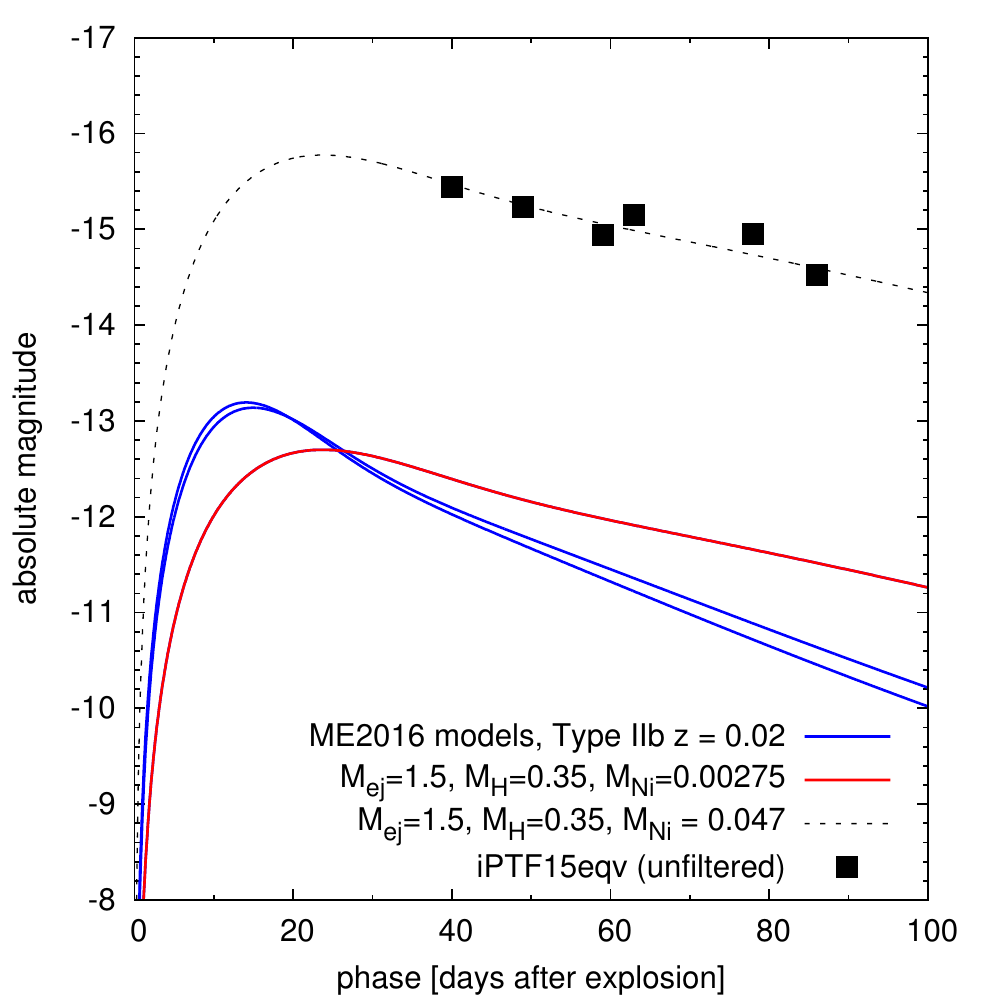}

\caption{The light curve of iPTF15eqv compared to bolometric light
  curves from models of ECSNe in close binary systems presented in
  \citet{ME16}. These models are for Type IIb explosions with Z =
  0.02, and progenitors with 0.35~M$_{\odot}$ of hydrogen at the time
  of explosion. We find reasonable consistency between iPTF15eqv and a
  model having a total ejecta mass of 1.5 M$_{\odot}$ if we scale the
  luminosity to mimic 0.047\,M$_{\odot}$ of nickel.}

\label{fig:me16}
\end{figure}

\subsection{Ca abundance and enrichment}

The potentially large calcium abundance of Ca-rich transients, coupled
with their large galactocentric distances, make them intriguing
candidate drivers of chemical evolution in galaxies and the
intergalactic medium \citep{Perets10,Kawabata10,Kasliwal12,MKK14}.
Recent works suggest that combinations of Type Ia and core-collapse
nucleosynthesis models under-produce the amount of calcium required by
X-ray observations of the intra-cluster medium
\citep{dePlaa07,Mernier16}, and that Ca-rich transients may play a
role in addressing this problem \citep{MKK14,Mernier16}.  Ca-rich
transients have also been invoked to explain metal-poor stars with an
extremely large over-abundance of Ca, such as SDSS J234723.64+010833.4
discovered in the outer halo of our galaxy \citep{Lai09}.

We derived O and Ca masses of $0.090_{-0.08}^{+0.11}$ and
0.006$\pm$0.002 M$_{\odot}$, respectively, for iPTF15eqv from our
nebular spectra. In contrast, estimates of chemical abundances of
SN\,2005E by \citet{Perets10} were 0.037 and 0.135 M$_{\odot}$ for O
and Ca, respectively. Taking into account that the [\ion{O}{1}] and
[\ion{Ca}{2}] emission lines of iPTF15eqv are $\sim 2.5$ and $\sim 20$
times more luminous than SN\,2005E, the O abundance estimates are
broadly consistent, whereas the Ca abundances are not. Some of the
inconsistency may be attributable to different model assumptions, but
additional discrepancies may arise from the different natures of the
explosions. Consequently, the Ca abundance yield of Ca-rich transients
may vary considerably across individual events.

Regardless of the exact nucleosynthetic yields, both analyses point to
a relatively large Ca/O abundance ratio, which is generally understood
to be sensitive to the progenitor mass in core-collapse SNe. Models of
supernova nucleosynthesis have routinely shown that larger masses of
Ca are made for lower mass progenitors
\citep{WH07,Nomoto13,Sukhbold16}. For example, theoretical models
predict that stars having main-sequence masses of 13 M$_{\odot}$ and
18 M$_{\odot}$ produce Ca/O ratios (by mass) of 0.025 and 0.005,
respectively \citep{Nomoto13}. Keeping with this trend, in order to
produce a relatively large Ca/O ratio of $\approx 0.07$ observed in
iPTF15eqv, the mass of its progenitor star would have been
$< 13$\,M$_{\odot}$ and thus approach the threshold for core collapse
($8\pm1$\,M$_{\odot}$; \citealt{Smartt09}).

\subsection{Electron capture supernova?}

Stars in the mass range $8-12$ M$_{\odot}$ can exhibit structural
peculiarities during their evolution that considerably affect the
supernova explosion dynamics if they undergo core collapse. Electron
captures on $^{24}$Mg and $^{20}$Ne in a degenerate O-Ne-Mg core of
mass ∼$\sim 1.37$\,M$_{\odot}$ can drive the core towards collapse in
what is referred to as an ``electron-capture supernova'' (ECSN;
\citealt{Nomoto84}). This progenitor scenario was invoked to explain
the Ca-rich SN\,2005cz \citep{Kawabata10}. The winds of such stars are
not strong enough to remove hydrogen envelopes, which led
\citet{Kawabata10} to suggest that binary interaction via Roche-lobe
overflow or common envelope ejection could explain why no conspicuous
hydrogen was present in their optical spectra of SN\,2005cz. A similar
progenitor scenario could apply to iPTF15eqv.

Generally, numerical simulations of ECSNe explode with kinetic
energies of $\sim 10^{50}$ erg and synthesize $ \sim 10^{-3}$
M$_{\odot}$ of radioactive $^{56}$Ni (e.g.,
\citealt{Kitaura06}). These values are each approximately an order of
magnitude less than those we infer for iPTF15eqv
(section~\ref{sec:lc}). However, there are many uncertainties with the
explosion processes of ECSNe (see, e.g.,
\citealt{Jones13,Woosley15}). Perhaps the strongest signature of an
ECSN is that it likely leaves behind a very special abundance pattern
in its ejecta (see the recent review by \citealt{Mueller16}), which
may be abundant in isotopes of calcium \citep{Wanajo13}.

\citet{ME16} investigated the expected rates and bolometric light
curve properties of stripped-envelope ECSNe using stellar models from
the Binary Population and Spectral Synthesis (BPASS)
code\footnote{http://bpass.auckland.ac.nz}
\citep{Eldridge08,Eldridge11,Stanway16}. We compared the light curve
of iPTF15eqv to the Type IIb, Z=0.020 models and were unable to find a
strict match. However, the \citet{ME16} models are fixed at
$2.5 \times 10^{-3}$ M$_{\odot}$ of $^{56}$Ni, which mainly changes
the peak luminosity. By scaling the luminosity of a model having a
total ejecta mass of 1.5 M$_{\odot}$ and hydrogen envelope mass of
0.35 M$_{\odot}$, reasonable agreement is possible if M$_{\rm Ni}$ is
allowed to range between $0.045-0.055$~M$_{\odot}$
(Figure~\ref{fig:me16}). The range in mass of our fit is due to
uncertainty in the explosion date. These results are consistent with
our independent derivation of M$_{\rm Ni}$ in section~\ref{sec:lc}.

Notably, binary systems of stars in the mass range that develop to
ECSNe can significantly extend the delay time distribution of
core-collapse SNe. For single stars the maximum lifetime is
approximately $\sim 50$ Myr. However, the majority of binary
progenitors in the \citet{ME16} models have lifetimes of 30 to
100\,Myr, and the oldest lifetime is 200\,Myr. A population synthesis
study conducted by \citet{Zapartas17} to compute the delay time
distribution of core collapse SNe also found that a significant
fraction can occur 50-200 Myr after birth, well after all massive
single stars have already exploded. These late events originate mostly
from binary systems with both stars of mass 4-8 M$_{\odot}$.

\subsection{Ultra-stripped supernova?}

Ultra-stripped SNe are core-collapse SNe whose ejecta masses are only
$\sim 0.1~M_\odot$ \citep{Tauris13,Tauris15}. The small ejecta masses
result from the extreme stripping of their progenitors due to the
existence of compact companion stars \citep{Tauris13,Tauris15}. The
ultra-stripped SNe are found to explode successfully with the
neutrino-driven mechanism, with explosion energies of
$\sim 10^{50}~\mathrm{erg}$ and $^{56}$Ni masses of
$\sim 0.01~M_\odot$ \citep{Suwa15}.  \citet{Moriya16} found that light
curves and spectra of some Ca-rich transients match well to those
predicted by ultra-stripped SNe, suggesting some Ca-rich transients
may be ultra-stripped SNe.

Ultra-stripped SN light curve models currently available tend to
decline faster than that of iPTF15eqv
\citep{Tauris13,Moriya16}. However, the current light curve models are
only available for progenitors with ejecta masses below $0.2~M_\odot$
\citep{Tauris13,Moriya16}. There exist ultra-stripped SNe with larger
ejecta masses \citep{Tauris15} and they can have slower decline rates,
possibly matching iPTF15eqv. One problem applying the ultra-stripped
model to iPTF15eqv is that iPTF15eqv likely exhibits hydrogen in our
late-time spectra (sections \ref{sec:identify} and \ref{sec:hydrogen};
Figure~\ref{fig:sn2011dh}). Ultra-stripped SN progenitors lack
hydrogen because of the extreme stripping, and their companions are
compact stars that do not have hydrogen. Therefore, no hydrogen is
expected to exist in the progenitor system. We consider an
ultra-stripped SN origin unlikely for iPTF15eqv, although further work
on this progenitor scenario is needed.

\begin{figure}

\includegraphics[width=\linewidth]{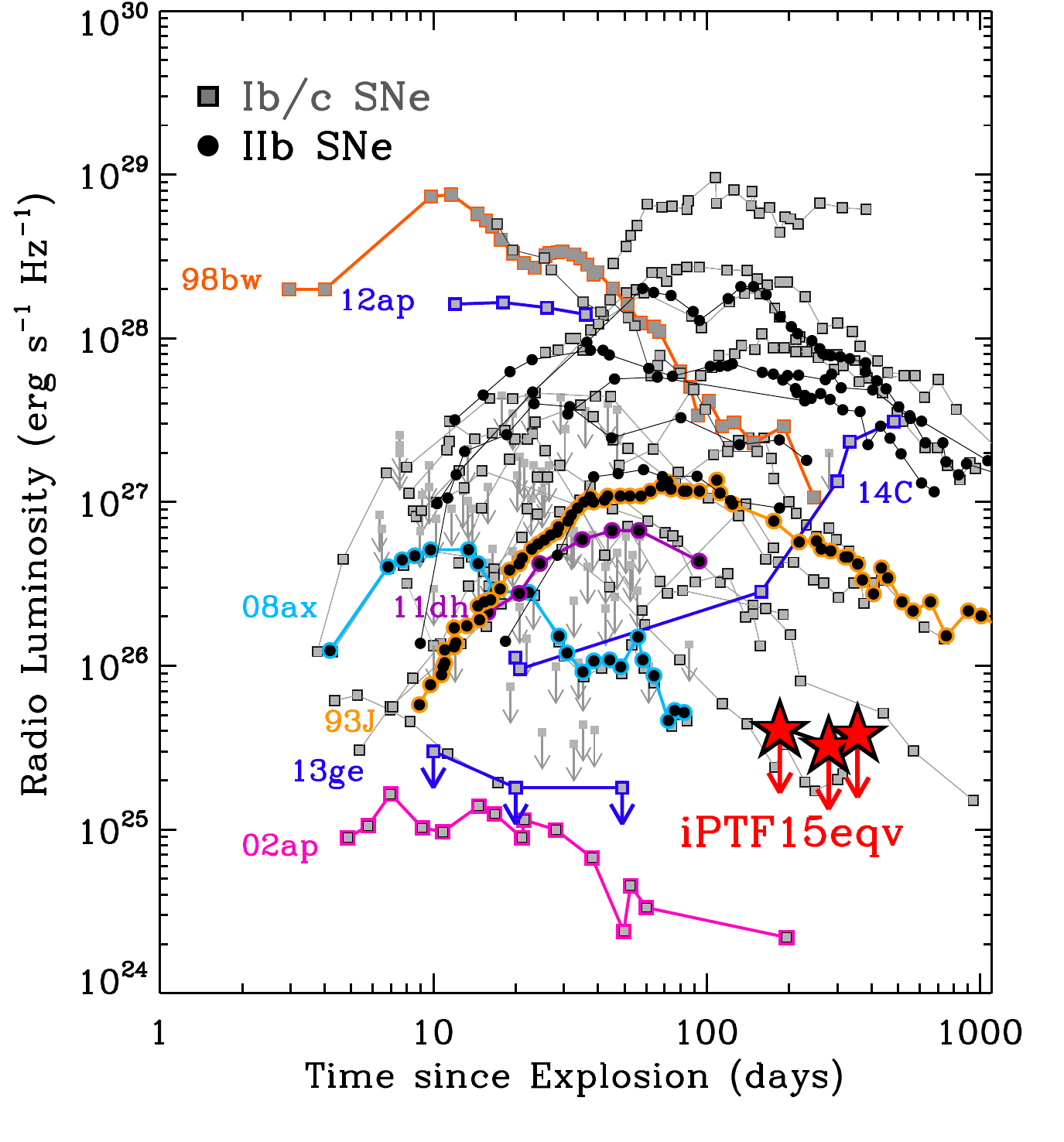}

\caption{Upper limits on radio emission from iPTF15eqv compared to the
  radio light curves of SNe IIb and Ib/c. Data were originally
  published in \citet{Margutti13} and references therein,
  \citet{Milisavljevic13} and references therein, \citet{Bufano14},
  \citet{Kamble16}, \citet{Kamble14}, \citet{Drout16}, and
  \citet{Margutti17}.}

\label{fig:radio-compare}
\end{figure}

\subsection{Mass loss environment}

Mass loss is a key process in stellar evolution, and thus is a
powerful probe of the progenitor systems of SNe
\citep{Puls08,Smith14ARAA}. Observations of SNe at radio and X-ray
wavelengths can reconstruct the mass loss history of its progenitor
system as the SN blast wave passes through and interacts with
surrounding material \citep{Chevalier82,Chevalier98}. To date no Type
Ia supernova has been detected at X-ray or radio wavelengths
\citep{Margutti12,Margutti14J,Chomiuk16}, suggesting a relatively
clean environment with densities below $n < 3$ cm$^{-3}$ at
$R \sim 10^{16}$\,cm and progenitor system mass loss rates of
$\dot{M} < 10^{-9}\, \frac{M_{\odot}\;\rm yr^{-1}}{100\; \rm
  km\;s^{-1}}$.  In contrast, core-collapse SNe exhibit a wide range
of X-ray and radio properties that reflect the diversity of their
environments that have been sculpted by the progenitor system's mass
loss
\citep{Weiler02,Kotak06,Soderberg06a,Chevalier10,Wellons12,Margutti13,Margutti17}.

\citet{Chomiuk16} examined radio upper limits for Ca-rich transients
in the context of Type Ia thermonuclear WD explosions. Assuming a
total ejecta mass of 0.3 M$_{\odot}$ and kinetic energy of
$4 \times 10^{50}$\,erg, they constrain
$\dot{M} < 10^{-5}\, \rm M_{\odot} yr^{-1}$ for $v_{w} = 100$ \kms\
for most objects of their sample. This allows them to rule out the
presence of tidal tails \citep{RK13}, in the case where stripping of
the He white dwarf occurred a few years before SN explosion, as well
as the presence of strong accretion-powered outflows or winds
\citep{Hachisu99}, as might be expected if a disrupted white dwarf is
accreted.

In light of the overlap in properties between iPTF15eqv and
core-collapse SNe (section \ref{sec:late-time}), we compared our VLA
upper limits on radio emission from iPTF15eqv to the radio light
curves of SNe IIb and Ib/c (Figure~\ref{fig:radio-compare}). Our
observations rule out environments similar to those observed around
the Type IIb SN\,1993J and SN\,2011dh, which exhibited radio light
curves that are more luminous than iPTF15eqv and were estimated to
have mass loss rates ranging between
$\dot{M} \approx 2.4 \times 10^{-5}$ M$_{\odot}$ yr$^{-1}$ ($v_w = 10$
\kms) and $\dot{M} \approx 6\times 10^{-5}$\,M$_{\odot}$ yr$^{-1}$
($v_w = 1000$ \kms), respectively. The environments of other Type IIb
such as SN\,2008ax with $\dot{M} = (9 \pm 3) \times 10^{-6}$
M$_{\odot}$ yr$^{-1}$ (for $v_w = 10$ \kms) are not ruled out
\citep{Roming09}. The difference in mass loss rates of SNe IIb may
reflect the size and evolutionary status of the progenitor star
\citep{Chevalier10,Kamble16}. Many SNe Ib/c such as SN\,2013ge
\citep{Drout16} and SN\,2002ap \citep{Berger02} with relatively clean
environments are not ruled out, but late-interacting SNe Ib/c such as
SN\,2014C are ruled out \citep{Milisavljevic15-14C,Margutti17}.

Our deep non-detections up to $\sim$ 400 days imply a low-density
environment ($n \la 0.1$ cm$^{-3}$) for the progenitor of iPTF15eqv
out to $R \sim 10^{17}$ cm. Using a
typical forward shock velocity of $\approx 3 \times 10^4$ \kms\ for
core-collapse SNe, we estimate that the progenitor system of iPTF15eqv
did not undergo any enhanced mass loss or eruptive activity within $\sim
300-1000$ ($v_w/100$ \kms) yr of the SN.

\section{Conclusions}
\label{sec:conclusions}

We have presented multi-wavelength observations and analysis of the
supernova explosion iPTF15eqv in NGC\,3430. We find that iPTF15eqv has
a combination of properties that bridge those observed in Ca-rich
transients and SNe Ib/c. Perhaps most revealing about iPTF15eqv is its
large [\ion{Ca}{2}]/[\ion{O}{1}] $\approx 10$ emission line ratio,
which is among the highest encountered among Ca-rich events, and clear
Type Ib/c and Type II spectroscopic signatures (Figure~\ref{fig:haha})
seen in our optical and NIR data $\ga 200$ days post-explosion. This
discovery is significant because discussions of Type Ia SNe often
include Ca-rich transients (e.g., \citealt{Parrent14,Chomiuk16}), and
our work establishes that an unknown percentage of SNe that have been
considered to be Ca-rich transients may be associated with massive
star explosions.

Our data allow us to place many constraints on the chemical abundance
yields of the explosion and the nature of the progenitor system of
iPTF15eqv:

\begin{enumerate}

\item We estimate M$_{\rm ej} \approx 2-4$\,M$_{\odot}$, M$_{Ni}$
  $\approx 0.04 - 0.07$ M$_{\odot}$, and E$_{k}$
  $\approx (0.8-2) \times 10^{51}$\,erg, assuming an explosion date
  sometime between 2015 July 24 and September 12.

\item We derive an O:Ca abundance ratio (by number) of 39$\pm$24, S:Ca
  abundance ratio of 29$\pm$7, and Si:S abundance ratio of $\sim 1$.

\item Upper limits on possible radio emission from iPTF15eqv made with
  the VLA suggest a low density environment
  ($n \la$ 0.1 cm$^{-3}$) within a radius of $\sim 10^{17}$ cm
  of the explosion site.

\item We do not observe X-ray emission that would be associated with
  accretion onto a BH to an upper limit of
  $5.3\times 10^{38}\,\rm{erg\,s^{-1}}$, which constrains the
  possibility of any BH to a mass $\lesssim 100 M_\odot$.

\item The leading candidate progenitor system is an H-poor star of
  mass $\la 10$\,M$_{\odot}$ that interacted with a nearby companion
  star. The chemical abundances and ejecta kinematics inferred from
  our late-time spectra suggest that the explosion may have been an
  ECSN.

\end{enumerate}


iPTF15eqv forces a sober reconsideration of its ``Ca-rich transient''
classification. On the one hand, without early epoch photometry and
spectra to constrain its light curve and specific SN classification,
iPTF15eqv does not strictly fit the criteria of a ``SN\,2005E-like''
\citep{Perets10} or ``Ca-rich gap transient''
\citep{Kasliwal12}. However, iPTF15eqv does exhibit strong calcium
line emission seen in other SNe having incomplete spectral and
photometric characterization (including 2003H, 2003dr, 2003dg, 2001co,
2000ds) that have been investigated as Ca-rich transients. It is
noteworthy that objects considered to be Ca-rich transients thus far
all share spectral evolution that truly distinguishes them from other
SNe (see Figures~\ref{fig:ca-rich-compare} and \ref{fig:CaIIvOI}),
which is an indication that the observational classification is
meaningful.

As data improve in quality and wavelength span (particularly in the
NIR; see section \ref{sec:identify}) the ability to define a further
subdivision in the class may be achievable, as was done in the Type I
classification of \citet{Minkowski41} that was later split to Type Ia
(thermonuclear explosions of WDs) and Type Ib/c (massive star core
collapses). However, reconciling how two radically different
progenitor channels can exhibit similar emission properties at nebular
epochs when emission is from interior ejecta will be challenging
(Figure~\ref{fig:haha}).

The relatively long delay-time distribution and distinguishable
abundance patterns of ECSNe from binary systems make them an
attractive progenitor system for at least some Ca-rich transients.  It
is important to note, however, that alternative models not immediately
applicable to iPTF15eqv, including ``ultra-stripped'' SNe and violent
disruptions of WDs, successfully reproduce other observed properties
of Ca-rich transients. Additional theoretical work that can provide
distinct signatures of the various progenitor scenarios is needed.

In the case of iPTF15eqv, high-resolution imaging obtained after the
SN has faded can map out the surrounding stellar environment and be
used to investigate whether a massive star association is nearby. More
generally, NIR spectra at nebular epochs proved to be critical for our
line identifications and diagnostics, and it is hoped that future
Ca-rich transients can be observed and analyzed in this way. Perhaps
the most pertinent issue for Ca-rich transients is to quantify their
calcium abundances more accurately.  Given that present estimates of
their yields vary across orders of magnitude, more stringent
constraints on their progenitor systems and improved models of their
late-time emissions are needed if a precise understanding of the
chemical enrichment of galaxies and the intracluster medium is to be
achieved.

\medskip

\acknowledgments We thank the referee H.\ Perets for a critical
reading of our manuscript and for providing many suggestions that
improved its content and presentation.  Observations reported here
were obtained at the MMT Observatory, a joint facility of the
Smithsonian Institution and the University of Arizona. Based on
observations made by the {\it Chandra X-ray Observatory} under program
GO 16500534, PI Margutti, observations ID 16821. The National Radio
Astronomy Observatory is a facility of the National Science Foundation
operated under cooperative agreement by Associated Universities, Inc.
This paper made use of the Weizmann interactive supernova data
repository (WISEREP) - http://wiserep.weizmann.ac.il - and the Open
Supernova Catalog - http://sne.space. K.\ Kawabata kindly provided
data of SN\,2005cz.  T.J.M.\ is supported by the Grant-in-Aid for
Research Activity Start-up of the Japan Society for the Promotion of
Science (16H07413).  R.M.\ gratefully acknowledges support from the
Research Corporation for Science Advancement. The authors acknowledge
the Telescope Data Center of the Smithsonian Astrophysical Observatory
for performing MMIRS data reduction. I.C.\ acknowledges partial
support by the Russian–French PICS International Laboratory program
(no. 6590) co-funded by the RFBR (project 15-52-15050), by the RFBR
project 15-32-21062 and the presidential grant MD-7355.2015.2. This
work is based in part on observations obtained at the MDM Observatory,
operated by Dartmouth College, Columbia University, Ohio State
University, Ohio University, and the University of Michigan.

\facility{MMT (Blue Channel spectrograph), MMT (MMTCam), MMT (MMIRS),
  Hiltner (OSMOS), EVLA, CXO}

\bibliographystyle{yahapj}

\clearpage
\newpage

\begin{deluxetable*}{lcccccc}
\tablecolumns{7}
\tablecaption{Log of optical photometry}
\tablehead{ \colhead{Date}    	    &
                   \colhead{Date}          &
                   \colhead{Phase\tablenotemark{*}} &
                   \colhead{Telescope} &
                   \colhead{Filter} &
                   \colhead{Magnitude} &
                   \colhead{$\sigma$} \\
                   \colhead{(UT)}       &
                    \colhead{(MJD)}    &
                    \colhead{(days)}    &
                    \colhead{+ Instr.} &
                    \colhead{}     &
                    \colhead{}      &
                    \colhead{}}
\startdata
2015 Sep 27  & 57292.81 &  0   & Itagaki Observatory & Clear & 
16.97 & 0.10\\
2015 Oct 06  & 57301.80 &   9    & Itagaki Observatory & Clear &
17.17 & 0.05\\
2015 Oct 16 & 57311.77 &   19     & Itagaki Observatory & Clear & 
17.47 & 0.11\\
2015 Oct 20 & 57315.76 &    23    & Itagaki Observatory & Clear &
17.26 & 0.08\\
2015 Nov 04 & 57330.80 &   38    & Itagaki Observatory & Clear &
17.46 & 0.14\\
2015 Nov 12 & 57338.82 &   46    & Itagaki Observatory & Clear &
17.88 & 0.09\\
2015 Dec 12 & 57368.72 &  76     & Itagaki Observatory & Clear &
18.34 & 0.10\\
2015 Dec 18 & 57374.79 & 82  & Itagaki Observatory & Clear &
18.46 & 0.17\\
2016 Feb 06 & 57425.04 &  132   & MMT+MMTCam & $r'$ & 19.70 &
0.06\\
2016 Feb 06 & 57425.05 & 132    & MMT+MMTCam & $i'$ & 18.26 &
0.08\\
2016 Mar 11 & 57458.61 & 166   & MMT+MMTCam & $r'$ & 20.35 &
0.10\\
2016 Mar 11 & 57458.62 &  166  & MMT+MMTCam & $i'$ & 18.90 &
0.08\\ 
2016 Jun 08 & 57547.65 & 255   & MMT+MMTCam & $g'$ & 22.12 &
0.09\\ 
2016 Jun 08 & 57547.65 &  255  & MMT+MMTCam & $r'$ & 21.76 &
0.12\\
2016 Jun 08 & 57547.66 & 255   & MMT+MMTCam & $i'$ & 20.90 &
0.09
\enddata
\label{tab:photometry}
\tablenotetext{*}{Phase is with respect to date of discovery on 2015 Sep
  27.8 (MJD 57292.81).}
  \tablecomments{\textbf{Note:} Unfiltered photometry should be interpreted as AB magnitudes.}
\end{deluxetable*}

\begin{deluxetable*}{llclcc}
\tablecolumns{6}
\tablecaption{Log of optical spectroscopy}
\tablehead{ \colhead{Date}    	    &
                   \colhead{Date}          &
                   \colhead{Phase\tablenotemark{*}} &
                   \colhead{Telescope} &
                   \colhead{Range} &
                   \colhead{Resolution} \\
                   \colhead{(UT)}       &
                    \colhead{(MJD)}    &
                    \colhead{(days)}    &
                    \colhead{+ Instr.} &
                    \colhead{(\AA)}     &
                    \colhead{(FWHM in \AA)}}
\startdata
2015 Dec 31.5  & 57387.5 & 95 & MDM\,2.4m+OSMOS & $5350-10500$ & 6\\
2016 Jan 18.3  & 57405.3   & 112 & MMT+Blue Channel & $3360-8580$ & 6\\
2016 Feb 15.3 & 57433.3   & 140 & MMT+Blue Channel & $3360-8580$ & 6\\
2016 Mar 02.4 & 57450.4   & 158 & MDM\,2.4m+OSMOS & $5350-10500$ & 6\\
2016 Apr 14.2 & 57492.2   & 199 & MMT+Blue Channel & $3360-8580$ & 6\\
2016 May $7-9$ & 57517.3 & 225 & MDM\,2.4m+OSMOS & $5350-10500$ & 6
\enddata
\label{tab:spectra}
\tablenotetext{*}{Phase is with respect to date of discovery 2015 Sep
  27.8 (MJD 57292.8).}
\end{deluxetable*}

\begin{deluxetable*}{ccccc}
\tablecolumns{5}
\tablecaption{Log of VLA observations}
\tablehead{ \colhead{Date}    	    &
                   \colhead{Date}          &
                   \colhead{Frequency} &
                   \colhead{$3 \times \sigma$ RMS} &
                   \colhead{VLA} \\
                   \colhead{(UT)}       &
                    \colhead{(MJD)}    &
                    \colhead{(GHz)}    &
                    \colhead{($\mu$Jy)}     &
                    \colhead{Configuration}}
\startdata
2015 Dec 24.57	&        57380.57	&	4.9     &	$<$	318    &	D    	\\   
2015 Dec 24.57	&        57380.57   &	7.1     &	$<$	543    &	D      \\
2016 Jan 05.60	&        57392.60   &	13.1   &	$<$	45    &	DnC	\\
2016 Jan 05.60	&        57392.60   &	16.0   &	$<$	42    &	DnC	\\
2016 Feb 18.48	&        57436.48   &	8.6     &	$<$	42    &	C	\\
2016 Feb 18.48	&        57436.48   &	11.0   &	$<$	43    &	C	\\
2016 May 22.22	&        57530.22   &	8.6     &	$<$	33    &	B	\\
2016 May 22.22	&        57530.22   &	11.0   &	$<$	37    &	B	\\
2016 Aug 03.08  &      	57603.08    &    4.9   &    $<$ 67    & B	\\   
2016 Aug 03.08  &       57603.08    &    7.1   &    $<$ 40    & B   \\
2016 Aug 03.08  &       57603.08    &    8.6   &    $<$ 38    & B   \\ 
2016 Aug 03.08  &       57603.08    &    11.0  &    $<$ 40    & B   
\enddata
\label{tab:vla}
\end{deluxetable*}

\begin{deluxetable*}{lrcrr}
\tablecolumns{5}
\tablecaption{[\ion{Ca}{2}]/[\ion{O}{1}] ratio of SN Ib/c and Ca-rich transients}
\tablehead{\colhead{Object} & 
                  \colhead{[\ion{Ca}{2}]/[\ion{O}{1}]} & 
                  \colhead{Phase}  &
                  \colhead{Type} & 
                  \colhead{Ref.} \\
                  \colhead{} &
                   \colhead{} &
                   \colhead{(days post-maximum)} &
                   \colhead{} &
                   \colhead{}}
\startdata
iPTF15eqv    & 11.3 &95\tablenotemark{$^{*}$}   & Ca-rich & This paper\\
iPTF15eqv    & 9.5   & 199\tablenotemark{$^{*}$} & Ca-rich & This paper\\
SN2005cz    & 7.5   & 179         & Ca-rich & \citet{Kawabata10}\\
PTF10iuv      & 6.8   & 87          & Ca-rich & \citet{Kasliwal12}\\
PTF11kmb    & 4.6   & 100        & Ca-rich & \citet{Foley15}\\
SN\,2005E    & 7.3  & 62          & Ca-rich   & \citet{Perets10}\\
SN\,2012hn  & 2.5  & 149        & Ca-rich  & \citet{Valenti14}\\
PTF11bij       & 4.4  & 45\tablenotemark{$^{*}$}  & Ca-rich & \citet{Kasliwal12}\\
SN\,2003dr  & 9.6  & 76\tablenotemark{$^{*}$}  & Ca-rich & \citet{Kasliwal12}\\
SN\,1999em & 4.2  & 313    & IIP & \citet{Leonard02}\\
SN\,2004dj & 1.8  & 252    & IIP & OSC\\
SN\,2004et & 3.0  & 301    & IIP & \citet{Sahu06}\\
SN\,2004et & 3.3  & 391    & IIP & \citet{Sahu06}\\
SN\,2012aw & 1.4  & 379    & IIP & \citet{Jerkstrand14}\\
SN\,2008ax & 1.39 & 100    & IIb & \citet{Milisavljevic10}\\
SN\,2008ax & 0.75 & 307    & IIb & \citet{Milisavljevic10}\\
SN\,2011dh & 1.31 & 152    & IIb & \citet{Ergon15}\\
SN\,2011dh & 0.96 & 272    & IIb & \citet{Ergon15}\\
iPTF13bvn  & 1.5  & 290    & Ib  & \citet{Kuncarayakti15}\\
SN\,2008D  & 0.76 & 363    & Ib  & \citet{Tanaka09}\\
SN\,2013ge & 0.39 & 159    & Ib/c & \citet{Drout16}\\
SN\,2013ge & 0.36 & 363    & Ib/c &  \citet{Drout16}\\
SN\,1994I  & 1.22 & 103    &  Ic & \citet{Flipper95}\\
SN\,1994I  & 0.98 & 153    & Ic  & \citet{Flipper95}\\
SN\,2012ap & 1.78 & 218    & Ic-bl & \citet{Milisavljevic15}\\
SN\,2012ap & 0.97 & 272    & Ic-bl & \citet{Milisavljevic15}\\
SN\,2002ap & 0.45 & 185    & Ic-bl & \citet{Foley03}\\
SN\,1998bw & 0.43 & 376    & Ic-bl & \citet{Patat01}
\enddata
\tablenotetext{*}{Phase is with respect to date of discovery.}
\label{tab:CaIIvOI}
\end{deluxetable*}

\begin{deluxetable*}{llrrrc}
\tablecolumns{6}
\tablecaption{Host galaxy properties of Ca-rich transients}
\tablehead{\colhead{SN} & 
                  \colhead{Host Galaxy} & 
                  \colhead{z}  & 
                  \colhead{$\rm{M}_g$} & 
                  \colhead{$\rm{M}_r$} & 
                  \colhead{$\log(\rm{O} / \rm{H})+12$ (LZC; PP04 N2 )}
}
\startdata
2000ds        &   NGC 2768   &   0.0045   &   -20.09      &   -21.00
&    8.71\\
2001co        &   NGC 5559   &   0.0172   &   -20.02      &   -20.69    &    8.71\\
2003dg        &   UGC 6934   &   0.0183   &   -19.46      &   -20.13    &    8.70\\
2005E         &   NGC 1032   &   0.0090   &   -19.69      &   -21.07    &    8.41\\
PTF11bij     &   IC 3956    &   0.0347   &   -20.49      &   -21.27    &    8.72\\
iPTF15eqv    &   NGC 3430   &   0.0053   &   -19.50      &   -20.08    &    8.67            
\enddata
\tablecomments{Photometry are SDSS~DR12 cModel magnitudes
  \citep{SDSSDR12} extinction-corrected for the Galaxy foreground
  \citep{Schlafly11}.  The oxygen abundance were derived with the
  LZC relation of \cite{Sanders13} using $g^{\prime}$ and $r^{\prime}$
  photometry and standardizing on the N2 diagnostic scale of
  \cite{PP04}.}

\label{tab:LZC}
\end{deluxetable*} 

\end{document}